\documentclass[sigplan,10pt]{acmart}

\usepackage{pifont}
\usepackage{newtxtext,newtxmath}
\usepackage{tabularx}
\usepackage{colortbl}
\usepackage{paralist}
\usepackage{mathtools}
\usepackage{multirow}
\usepackage{soul}
\sethlcolor{yellow}
\DeclareMathOperator*{\argmin}{argmin} 

\AtBeginDocument{%
  \providecommand\BibTeX{{%
    \normalfont B\kern-0.5em{\scshape i\kern-0.25em b}\kern-0.8em\TeX}}}

\acmYear{2025}\copyrightyear{2025}

\setcopyright{rightsretained}
\acmConference[EuroSys '25]{Twentieth European Conference on Computer Systems}{March 30--April 3, 2025}{Rotterdam, Netherlands}
\acmBooktitle{Twentieth European Conference on Computer Systems (EuroSys '25), March 30--April 3, 2025, Rotterdam, Netherlands}
\acmDOI{10.1145/3689031.3696072}
\acmISBN{979-8-4007-1196-1/25/03}

\usepackage[linesnumbered,ruled,vlined]{algorithm2e}
\newcommand{\sysname}{\textsc{HCache}}
\newcommand{\hidden}{\text{hidden states}}
\newcommand{\hcache}{\textsc{HCache}}
\newcommand{\kvcache}{\text{KV cache}}

\newcommand{\parabf}[1]{\medskip\noindent\textbf{#1}}

\begin{document}
\title{Fast State Restoration in LLM Serving with \sysname{}}
\author{Shiwei Gao}
\affiliation{%
  \institution{Tsinghua University}
}

\author{Youmin Chen}
\affiliation{%
  \institution{Tsinghua University}
}

\author{Jiwu Shu}
\authornote{Jiwu Shu is the corresponding author (shujw@tsinghua.edu.cn).}
\affiliation{%
  \institution{Tsinghua University}
}


\begin{abstract}
The growing complexity of LLM usage today, e.g., multi-round conversation and 
retrieval-augmented generation (RAG), makes contextual states (i.e., \kvcache{}) reusable across user requests. 
Given the capacity constraints of GPU memory, only a limited number of contexts can be cached on GPU for reusing. Existing inference systems typically evict part of the KV cache and restore it by 
recomputing it from the original tokens or offloading it to host storage for later retrieval, both of which introduce substantial computational or I/O overheads.

We propose \sysname{}, a novel LLM state restoration method. Its key idea is to restore LLM states from intermediate activations and thus utilize 
computational and I/O resources with low overhead. We enhance \hcache{} with two techniques,
including i) a \emph{bubble-free restoration scheduler} that integrates resource-complementary methods 
to optimize the balance between computation 
and IO tasks; and ii) a \emph{chunk-based storage manager} to address the layout 
mismatch issue (i.e., layer-before-token saving versus token-before-layer restoration). 
Our evaluations, conducted using real-world tasks, 
show that \sysname{} reduces the TTFT by up to 1.93$\times$ 
compared to KV offload while consuming 1.92-2.40$\times$ less storage space;
compared to token recomputation, \sysname{} achieves up to 5.73$\times$ reduction in TTFT.

\end{abstract}




\settopmatter{printfolios=true}
\maketitle

\section{Introduction}
\begin{figure}[t!]
    \centering
    \includegraphics[width=0.9\linewidth]{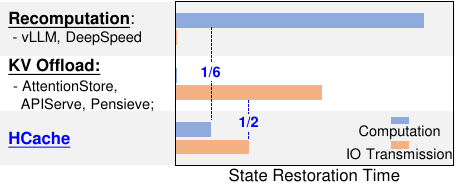}
    \vspace{-0.1in}
    \caption{State Restoration Method Comparison. \emph{\small{
    Recomputation: compute \kvcache{} from history tokens when reused;
    \kvcache{} offload: save \kvcache{} at host storage and fetch them to GPU memory when reused;
    \hcache{} saves 6$\times$ computational and 2$\times$ IO resources.
    }}}
    \vspace{-0.2in}
    \label{fig:intro}
\end{figure}
Large Language Models (LLMs)~\cite{chatgpt,touvron2023llama,gemmateam2024gemma,javaheripi2023phi} have demonstrated substantial potential across a wide array of applications, such as content generation~\cite{cao2023comprehensive,zhao2023more}, code comprehension~\cite{nam2024using,chen2021evaluating}, and the development of intelligent agents~\cite{Wang_2024,xi2023rise}. 
LLMs commonly serve user queries in two phases, including a \emph{prefill phase} to process a user's prompt and 
a \emph{generation phase} that sequentially generates new tokens autoregressively.
During both phases, intermediate states (a.k.a., \kvcache{}) are generated and cached for each token to avoid their recomputation.

Many existing LLM serving systems are stateless among user requests; the \kvcache{} is discarded once the request is finished. Some recent techniques aim at
optimizing individual requests, such as request batching~\cite{yu2022orca}, GPU memory management~\cite{stoica2023vllm}, 
and request scheduling~\cite{deepspeed2024fastgen,agrawal2024taming,wu2023fast}, among others. However, the increasing complexity of LLM usage today 
makes user requests depend on historical contexts.
For example, in multi-round conversational chatbots, the historical dialogue, including a user's 
prompts and the LLM's outputs, is accumulated as contextual states to enhance the LLM's understanding 
of the conversation background. In typical LLM chatbots such as Claude~\cite{claude}, the length of history dialogues 
can reach 200K~\cite{claudelong}.  There are also long-context applications, such as Retrieval-Augmented Generation (RAG).
RAG applications search domain-specific knowledge related to user requests and use it to help LLMs generate more accurate answers and alleviate the hallucination problem~\cite{huang2023survey,tonmoy2024comprehensive}. 
We call these applications as \emph{stateful} LLMs.

Yet, stateful LLMs raise new challenges for managing the states of their historical contexts. 
When serving new user requests, simply prefilling the prompt with historical contexts 
(i.e., token recomputation) will significantly increase computing overhead, potentially causing response time 
to multiply by tens of times. Recent work proposes directly reusing the \kvcache{} among user requests~\cite{gim2024prompt,zheng2023efficiently,ye2024chunkattention}. 
Nevertheless, our analysis of two popular LLM traces, i.e., ShareGPT4~\cite{sharegpt4} and L-Eval~\cite{an2023leval}, 
shows that a single A100-40GB GPU can keep up to 7-20 multi-round conversations or 1-3 long contexts. 
As a result, large-volume \kvcache{}s of historical contexts must be offloaded to host storage (i.e., 
KV offload~\cite{zuo2024as,yu2023stateful,yiying2024apiserve}), 
which again introduces high latencies when loading \kvcache{}s from host storage to GPU memory.

We observe that both token recomputation and KV offload adhere to the standard LLM inference process,  maintaining the LLM forward pass or \kvcache{} \textit{as is} to manage and restore historical contextual states. Consequently, these approaches occupy the 
extreme ends of the design spectrum, solely utilizing GPU computational or I/O  
capabilities, thus leaving system resources underutilized.
In this paper, we introduce \hcache{}, a novel LLM state restoration method 
to break this conventional wisdom by using GPU computational and IO resources simultaneously in a low overhead manner compared with existing methods (see Figure~\ref{fig:intro}).

At the core of \sysname{} is to exploit LLM's \textit{intrinsic} structure characteristic -- 
regaining the \kvcache{} by computing it from rather smaller intermediate activations -- for fast state restoration. 
Fortunately, the \hidden{} between adjacent LLM layers have such good properties and can be used 
as an alternative to the \kvcache{}. 
Specifically, the \hidden{} are half the size of the \kvcache{}, thus reducing the IO transmission time 
by 2$\times$ compared to KV offload. Furthermore, recomputing \kvcache{} from \hidden{} 
instead of original tokens eliminates two computationally intensive modules in LLM, including 
the quadratic complexity attention mechanism and the feed-forward network (FFN); 
hence, \sysname{} reduces the computation cost by at least 6$\times$. 

Restoring the contextual states with \hcache{} 
involves two parts -- transmitting the \hidden{} to GPU memory and recomputing it into \kvcache{} via common GEMM operations. By applying pipeline techniques, the above two steps can be performed concurrently to utilize both the IO and computational resources, further reducing state restoration time.
As a whole, \hcache{} significantly reduces the computation and IO overhead and is faster 
than existing approaches on mainstream platforms.
However, fully leveraging the capabilities of \hcache{} presents two technical challenges and 
we design techniques tailored to each issue.

First, state restoration with \hcache{} involves both recomputation and IO transmission, whose completion times are 
not always the same under varying hardware setups, producing pipeline bubbles. The actual restoration 
speed is bounded by the slower one of the two steps. To address this problem, we introduce the 
bubble-free restoration scheduler, which incorporates a resource-complementary restoration method to 
eliminate pipeline bubbles. Second, \hidden{} are generated and saved autoregressively in a \emph{layer-before-token} order, 
but are fetched as a whole batch in a \emph{token-before-layer} order in restoration. 
This incurs mismatched storage visiting orders in saving and restoration, 
presenting a hard design trade-off in the storage format to manage \hidden{} effectively. 
A storage format optimized for one stage inevitably results in small and random IOs for the other stage. 
In \hcache{}, we introduce a chunk-based storage format tailored for fast state restoration since 
reducing state restoration time is our primary design goal. 

To mitigate the performance impact on the generation phase,
we use a two-stage saving strategy to store newly generated \hidden{}.

We implement \sysname{} by integrating it into an LLM serving system and evaluating it 
with varying models and real-world LLM traces. 
Evaluation results show that \sysname{} reduces the TTFT (Time To First Token) by up to 1.93$\times$ against 
KV offload, adding less than 4\% overhead on TBT (Time Between Token). Compared to token recomputation, 
\sysname{} reduces TTFT by up to 5.73$\times$. We also evaluated \sysname{} 
on platforms with various computing and transmission speeds. 
With different hardware configurations, \sysname{} improves the restoration speed by 1.33-2.66$\times$ 
against KV offload.

In summary, we make the following contributions:
\begin{compactitem}[\scriptsize{$\bullet$}]
\item We propose \hcache{}, a novel method to restore historical LLM states utilizing intermediate activations.
\item We introduce the bubble-free restoration scheduler and chunk-based storage management to enhance \hcache{}.
\item We evaluate \sysname{} on various hardware setups and show that it outperforms the state-of-the-art solutions such as AttentionStore~\cite{zuo2024as} and DeepSpeed-MII~\cite{MII}.
\end{compactitem}
\section{Background and Motivation}
\subsection{Transformer Architecture}
\label{bg::subsec::transformer}

\begin{figure}
    \centering
    \includegraphics[width=\linewidth]{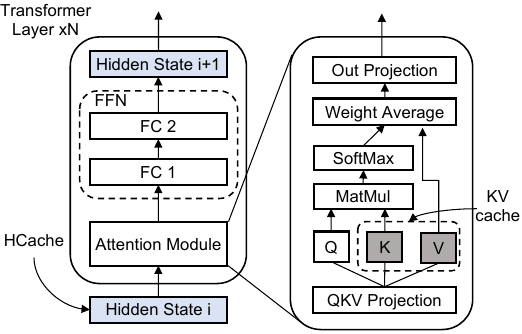}
    \vspace{-0.25in}
    \caption{Transformer Architecture. }
    \vspace{-0.2in}
     \label{fig:llmstructure}
\end{figure}

In this section, we first provide the background on the transformer architecture 
(\S\ref{bg::subsec::transformer}) and LLM serving systems (\S\ref{bg::subsec::llm}).
Then, we show the examples, characteristics, and overhead of stateful LLM 
applications in \S\ref{sec:bg:stateful} and \S\ref{bg::subssec::overhead}.

Most LLMs today adopt the transformer architecture as their building block. As shown in Figure~\ref{fig:llmstructure},
LLM has N repetitive transformer layers,
each comprising two major components: the attention module and the FFN (feed-forward network). 

An LLM forward pass is conducted with a batch of tokens. At the start of each layer, 
each token has a high-dimension representation, we call it the \emph{hidden states}. 
The first layer's \hidden{} come from the embedding table. The \hidden{} in other layers are the output of the previous transformer layer.
The attention module projects each token's hidden states to three different tensors: K, Q, and V. 
The model then uses the muti-head-attention (MHA)~\cite{vaswani2017attention} to interact information between tokens. 
Every token first computes its attention score by getting the softmaxed inner product of their $Q$ with the $K$ 
of other tokens. Next, the token computes the weighted average of $V$ of other tokens using their corresponding attention score. The weighted average is then multiplied by the out projection to get the final attention result.
The attention module in one layer can be summarized by the following equations:
\setlength{\abovedisplayskip}{3pt}
\setlength{\belowdisplayskip}{3pt}
\begin{equation*}
\begin{aligned}
Q_L^i, K_L^i, V_L^i &= W_L^{q, k, v}*H_L^i \\
\mathit{Attention}_L^i(q,k,v) &= \sum_{m=1}^{i-1} \frac{e^{Q_L^i*K_m}}{\sum_{n=1}^{i-1}e^{Q_L^i*K_n}}*V_i \\
Out_L^i &= W_L^o*\mathit{Attention}_L^i(q,k,v) 
\label{eq:attention}
\end{aligned}
\end{equation*}

FFN module consists of two linear projections with an activation function in between. 
The output of the FFN module (i.e., \hidden{}) is the input of the next transformer layer.
\begin{equation*}
\begin{aligned}
H_{L+1}^{i} = \mathit{FC2 }~(\mathit{ FC1 }~(Out_L^i)) \\
\end{aligned}
\end{equation*}

The above equations ignore less compute-intensive modules, including residual connection, 
layer norm, input embedding, and position embeddings. 
The internal architectures vary between different LLMs. We only discuss the common 
transformer structure in LLMs without loss of generality.

\kvcache{} is an important design to speed up existing LLM inference engines. 
In the attention module, the attention computation of a token depends on all previous 
KV values in the sequence. To avoid redundant KV recomputation, once a token's 
key and value are generated, they are stored in a dedicated storage area on GPU, i.e., the \kvcache{}. The generation for
the latter tokens of the sequence can visit \kvcache{} instead of recomputing KV values of history tokens.

\subsection{LLM Serving}
\label{bg::subsec::llm}
Once trained, LLMs can be deployed as a service to respond to user requests.
LLM inference systems process a user request in two phases -- 
\emph{prompt prefilling} and \emph{response generation}. 
During the prompt prefilling phase, LLM performs a transformer forward pass with input prompt tokens, 
generating the \kvcache{} for them. In the response generation phase, LLM generates new 
tokens autoregressively and puts the key-value tensors of the newly generated token 
into the \kvcache{}. One sequence's generation process continues until reaching an end-of-sequence (EOS) 
token or a maximum token budget.

Inference systems typically process user requests in batch to improve LLM serving efficiency. 
Instead of a stop-wait batching mechanism, which may incur high latency for early-arrived requests, 
most current LLMs use continuous batching~\cite{yu2022orca}. Specifically, 
the inference engine batches multiple requests in the generation phase at a fine-grained iteration level,
where new requests can be added (exited) to (from) a batch dynamically at each iteration. 
TTFT (Time To First Token) and TBT (Time Between Token) are important metrics describing the LLM serving quality. TTFT is the time it takes to generate the first token;
TBT represents the average time to generate a token for each request except for the first token.

\subsection{Stateful LLM Applications}
\label{sec:bg:stateful}
The basic usage of LLM assumes that LLM inference is \emph{stateless} -- user requests are independent and 
irrelevant to previous ones, and the \kvcache{} is discarded when the generation is finished. However, the increasing complexity of LLM usage today provides user requests with 
rich contextual contents, such as chat history and auxiliary documents, to help LLMs provide context-dependent, 
accurate, and informative responses. Hence, a request may rely on the context previously assimilated by the engine, making LLM inference a \emph{stateful} task. Below, we characterize different stateful LLM tasks.

\begin{figure}
    \centering
    \includegraphics[width=\linewidth]{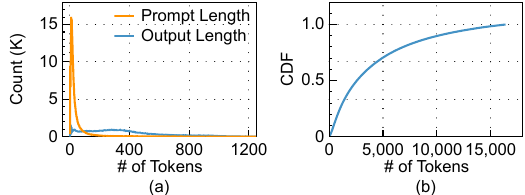}
    \vspace{-0.2in}
    \caption{Charactreistic of Multi-Round Conversation.}
    \vspace{-0.2in}
    \label{fig::moti::multiround}
\end{figure}

\parabf{Multi-round conversations.} 
Taking ChatGPT~\cite{chatgpt} as an example, a user starts a conversation by creating 
a ``New Chat'' session. In this session, the user keys in an instruction and waits for the response from ChatGPT; 
based on the response, the user can further send new questions to start a next-round conversation. 
To produce better answers in a multi-round conversation, the LLM generates a new response by considering both the 
tokens from all previous conversation rounds and the tokens from the current input.

A representative dataset of multi-round conversation is ShareGPT4~\cite{sharegpt4}, a trace of human conversations with GPT-4. 
Figure~\ref{fig::moti::multiround}a shows the average length of new prompt tokens and 
output tokens in one round. The average input length of each round is 66.8 tokens, and the output is 358.8 tokens. 
Though one round's input and output length is relatively short, they gradually accumulate in the conversation's history as 
the dialogue progresses. Figure~\ref{fig::moti::multiround}b depicts the CDF of the length of history tokens (truncated at 16K), and we can see that 
the length of half of the conversations is over 2.5K.

\parabf{Long-context applications.} 
Many LLM applications (e.g., LangChain~\cite{Chase_LangChain_2022}) also supplement 
user prompts with additional contexts to help LLM react better.

\begin{compactitem}[\scriptsize{$\bullet$}]
\item In contextual question and answer (Q\&A) applications, 
LLMs receive a long context text, such as an academic paper, law document, or textbook, and answer related user questions. 
For example, a user can upload a PDF of a language programming guide; after processing it, 
the LLM can answer the demo usage of different APIs.
\item Retrieval-Augmented Generation (RAG) application is also a popular workflow of LLMs~\cite{lewis2020retrieval,Liu_LlamaIndex_2022,Borgeaud2021ImprovingLM}. 
In response to user requests, RAG applications first search question-related content in their 
inner databases or on the Internet. Then, the retried documents and the user question 
are supplemented to the LLM. LLM uses the retrieved auxiliary knowledge to generate more accurate results for user questions. 
\end{compactitem}

\begin{table}[t!]
    \centering
    \arrayrulewidth=0.5pt
    \extrarowheight=1pt
    \setlength{\tabcolsep}{3mm}{
        \begin{tabular}{cccccc}
            \arrayrulecolor{black}\hline
            \arrayrulecolor{black}\hline
            \textbf{Tasktype} &\textbf{Context} & \textbf{Input} & \textbf{Output} \\ 
            \hline
            Paper Assistant & 10603.5 & 142.7 & 404.8\\ 
            GSM-100~\cite{cobbe2021training} & 5451.7 & 77.4 & 4.3 \\ 
            QuALITY~\cite{pang-etal-2022-quality} & 7053.9 & 92.4 & 19.2  \\
            \textbf{AVG. (20 sub-tasks)}& 16340.2 & 44.7 & 50.2 \\
            \arrayrulecolor{black}\hline
            \arrayrulecolor{black}\hline
        \end{tabular}
    }
    \caption{Statistics of the L-Eval~\cite{an2023leval} Dataset.}
    \vspace{-0.35in}
    \label{tab:moti:longcontext}
\end{table}
L-Eval~\cite{an2023leval} is an open-sourced dataset tailored for long-context LLM evaluation.
It contains 20 sub-tasks, including long context Q\&A, few-shot examples helped reasoning, and deducing the output of a piece of code. In Table~\ref{tab:moti:longcontext}, 
we present the statistics of three representative sub-tasks in L-Eval. 
These traces exhibit bimodal characteristics -- the average length of context contents 
can extend up to 16K. In contrast, the lengths of instructions and outputs typically remain below 100.

\subsection{State Restoration and Its Overhead}
\label{bg::subssec::overhead}
In stateful LLM inference, the \kvcache{} of shared tokens can be reused across different requests. Recent studies~\cite{rtp-llm,ye2024chunkattention,gim2024prompt} have explored caching and reusing the \kvcache{} between different requests on GPU to reduce the TTFT of user requests. However, the limited memory capacity of GPUs poses a significant challenge in retaining all necessary \kvcache{}. For instance, PagedAttention~\cite{stoica2023vllm}, a state-of-the-art \kvcache{} management technique, enables an A100-40G GPU to store up to 17K tokens with Llama2-13B and 48K tokens with Llama2-7B, which equates to only 1–3 extended contexts or 7–20 conversation sessions.

Due to the GPU memory limitation~\cite{xie2020kraken}, current inference systems often evict less frequently accessed \kvcache{} to free up space for more immediate contexts~\cite{zheng2023efficiently,stoica2023vllm}. When evicted \kvcache{} is needed again, it must be restored. We refer to this process as \emph{state restoration} in stateful LLM serving. Prior work has demonstrated that state restoration is a prevalent issue in inference systems. In multi-turn conversations, the interval between interactions often leads to a low GPU cache hit ratio~\cite{zuo2024as}, and in long context applications, the same context may be reused hours apart~\cite{liu2024cachegen}.

Two primary approaches exist for restoring LLM state. The first method is recomputation, which involves re-executing the prefill phase to regenerate the \kvcache{} from the original tokens~\cite{TensorRT-LLM,stoica2023vllm,MII,lightllm}. The second method is offloading the \kvcache{} to host memory or storage devices~\cite{zuo2024as,liu2024cachegen,jin2024ragcache,yiying2024apiserve}, with it being transmitted back to GPU memory when needed for subsequent requests.

\begin{figure}
    \centering
       \includegraphics[width=\linewidth]{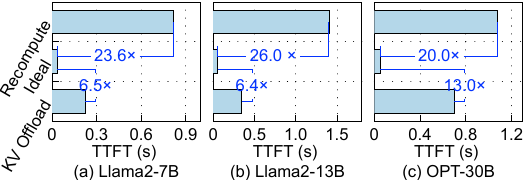}
    \vspace{-0.25in}
    \caption{Comparison of State Restoration Overhead.
    \emph{\small{
    We use the L-Eval trace; 
    Opt-30B runs on 4$\times$ A100-40G GPUs with tensor parallelism, while the rest two run on one A100; 
    4$\times$ PM9A3 SSDs are used as the storage backend to save offloaded \kvcache{}s.
    }}}
    \vspace{-0.2in}
    \label{fig:moti}
\end{figure}

Both state restoration methods result in substantial performance degradation compared to the ideal case where no restoration is needed. Our evaluation in Figure~\ref{fig:moti} shows that the TTFT for recomputation is 20.0-26.0$\times$ slower than the ideal case, while KV offloading is 6.5-13.0$\times$ slower.

A key point of the inefficiency of current state restoration methods relates to how they utilize the 
hardware resources. At one extreme, recomputation solely relies on GPU's computation capability to restore 
the \kvcache{} from tokens, whose computational complexity increases quadratically~\cite{peng2023rwkv} with the 
sequence's length. As the contrasting approach, KV offload solely relies on 
host storage and PCIe bus to load large-volume \kvcache{}, whose size is $10^5$ larger than 
that of history tokens in typical LLM models~\cite{liu2024cachegen}.
As a result, both approaches incur non-negligible overhead for \kvcache{} restoration.


\section{\hcache{}: Caching the Hidden States}
To accelerate the state restoration speed,
we aim to seek a new approach that \emph{simultaneously utilizes the computation capability and transmission bandwidth, 
while reducing unnecessary data transfers and recomputations}.
Our initial attempt is a 
simple combination of recomputation and KV offload. Specifically, the contextual tokens 
are divided into two parts, and we use recomputation and KV offload to restore them concurrently; 
hence, the computation and transmission resources can be utilized in parallel. 
However, this naive hybrid approach still keeps the LLM forward pass or KV cache as is, 
which does not fundamentally reduce the computation and IO overhead, resulting in suboptimal performance (see \S\ref{sec:eval:hybrid}).

To this end, we introduce \hcache{}, a novel state restoration method that efficiently 
utilizes both computational and I/O resources with reduced computing and transferring overhead.

\subsection{\hcache{} Overview}

\label{disign:hcache:overview}

Instead of directly restoring the \kvcache{}, 
the core idea behind \hcache{} is to utilize the LLM's intermediate 
activations (i.e., hidden states) at each transformer layer for state restoration. 

Recall from Figure~\ref{fig:llmstructure} that the \hidden{} represent the input to each transformer layer.
Within each transformer layer, the KV tensor pairs of each token are derived directly from the 
\hidden{} by performing the projection operation in the attention module. Specifically, 
we can use the following equations to restore the \kvcache{} for token $i$:
\begin{equation*}
    \begin{aligned}
    \label{eq:kv_restore}
    K_L^i &= W^k_L \cdot H_L^i \\
    V_L^i &= W^v_L\cdot H_L^i 
    \end{aligned}
\end{equation*}
where $H_L^i $ is the hidden state for a token $i$ at layer $L$, $K_L^i$ and $V_L^i$ represents 
the key and value tensors for token $i$ at that layer, and $W^k_L$ and $W^v_L$ are the 
linear projection parameters. 

An LLM serving system working with \hcache{} makes the following changes to reuse contextual contents.
First, \hidden{} should be saved for future reuse. Specifically, in multi-round conversations, 
whenever a layer's \hidden{} is generated in a LLM forward pass, 
it is dumped to the host storage concurrently with this layer's computation;
in RAG applications, \hidden{} can be generated and saved offline.
Upon new user requests arrive, the related \hcache{} is retrieved from the 
host storage into the GPU memory,
then, we can use the GEMM operator to restore the \kvcache{} from \hidden{} with the above equation. 

\begin{figure}
    \centering
    \includegraphics[width=\linewidth]{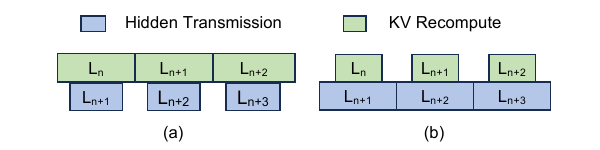}
    \vspace{-0.3in}
    \caption{ Example of Pipelined Restoration with \hcache{}.}
    \vspace{-0.15in}
    \label{fig:hcache:pipeline}
\end{figure}

Importantly, transmission and computation can be pipelined to utilize both the computation capability 
and transmission bandwidth at the same time. As shown in Figure~\ref{fig:hcache:pipeline}, 
the transmission of $L_{n+1}$ and recomputation from the \hidden{} to \kvcache{} of $L_{n}$ can be done concurrently.

Next, We theoretically analyze how caching \hidden{} can reduce both computational and data transfer volumes.

\subsection{Restoration Cost Analysis}
We compare \hcache{} with the two aforementioned methods: KV offload and recomputation. We analyze the restoration for one transformer layer that uses muti-head-attention (MHA).

\parabf{Restoration time with \hcache{}}. 
Restoration from \hidden{} involves two parts: \hidden{} transmission and 
recomputation from the \hidden{} to \kvcache{}. The time for transmitting \hidden{} is:
$$
IO_{hidden} = \frac{N_{seq} * D_{hidden}}{BW},
$$
where $N_{seq}$ indicates the number of tokens in the sequence, $D_{didden}$ is the LLM hidden dimension, and 
$BW$ is the bandwidth between hidden states' GPU and storage backend. 
The computation time\footnote{A multiply-add operator is regarded as 2 float point operations.} from \hidden{} to \kvcache{} is: 
$$
C_{hidden} = \frac{4 *N_{seq} * D_{hidden}* D_{hidden}}{FLOPS},
$$
where $FLOPS$ is the FLOPS of the GPU.
With the computation-transmission pipeline implemented in Figure~\ref{fig:hcache:pipeline},
the end-to-end restore time is dominated by the maximum of $IO_{hidden}$ and $C_{hidden}$. Hence,
the state restoration time of \hcache{} is:
$$
T_{hidden} = max(IO_{hidden},C_{hidden}).
$$

\parabf{Restoration time with KV offload}. The restoration process of KV offload only involves 
transferring the \kvcache{} from the host to GPU memory, whose time is:
$$
T_{kv} = IO_{KV} = \frac{2 *N_{seq} * D_{hidden}}{BW}.
$$

\parabf{Restoration time with recomputation.}
The restoration process of recomputation involves a full computation from history tokens to their \kvcache{}s. 
We ignore the time required to transfer tokens to GPU since their size is small.
\begin{align*}
C_{attn} &= \frac{8*N_{seq}* D_{hidden}^2  + N_{seq}^2*D_{hidden}}{FLOPS} \\
C_{ffn} &= \frac{16 * N_{seq}* D_{hidden}^2}{FLOPS} \\
T_{rec} &= C_{attn} + C_{ffn} + \epsilon \\
&= \frac{24 *N_{seq}* D_{hidden}^2  + N_{seq}^2*D_{hidden}}{FLOPS} + \epsilon
\end{align*}
Here $\epsilon$ indicates the remaining computation overhead, 
including normalization and residual connection, which are bounded by $O(N_{seq}*D_{hidden})$ and 
can be negligible. \\

\parabf{Comparison.}
\label{sec:hcache:compare}
For the I/O transmission part, the hidden state tensors have the same shape as keys and 
values in the \kvcache{}. Hence, the I/O time of transmitting \hidden{} is half that of directly transmitting the \kvcache{}. 
For the computation part. The relative speedup of \hcache{} compared with recomputation is: 
$$\frac{24 * N_{seq}* D_{hidden}^2+N_{seq}^2*D_{hidden}}{4 *N_{seq} * D_{hidden}^2}=6+\frac{N_{seq}}{4*D_{hidden}}.$$ 
Hence, the lower bound for the speed up is 6$\times$. 
The computational speedup is attributed to two factors. First, the FFN and attention modules 
are highly compute-intensive, reflected by the constant 24 in the above equation. 
In contrast, the projection of \hidden{} to the \kvcache{} is relatively lightweight, corresponding to a constant of 4.
This ensures a minimum 6$\times$ speed-up. Second, the quadratic complexity attention computation is the dominant component for long sequences but is unnecessary in restoring the hidden state. So \hcache{} scales linearly with history length.

\parabf{Conclusion.} As a whole, the transmission size of \hcache{} is always 2$\times$ less than 
that of transmitting the \kvcache{}, 
while the recomputation of \kvcache{} from \hidden{} is at least 6$\times$ faster than token recomputation.
This theoretical support makes \hcache{} beneficial on mainstream platforms (see \S\ref{sec:eval:sensi}).

\noindent Nevertheless, when employing \hcache{} to existing LLM serving systems, we still face two main challenges:

\label{sec:cache:challenge}
 
\parabf{C1: Pipeline bubbles.} As shown in Figure~\ref{fig:hcache:pipeline}, the computation and transmission of \hcache{} restoration 
do not always take the same time. Hence, the actual restoration speed is bounded by the slower one, which creates 
pipeline bubbles and wastes system resources. 
In extreme hardware configurations, where the IO speed far exceeds computation speeds, 
or vice versa, \hcache{} may not offer any benefits.

\begin{figure}
    \centering
    \includegraphics[width=\linewidth]{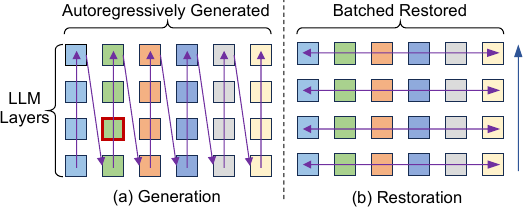}
    \vspace{-0.25in}
    \caption{State Generation and Restoration Order.
    \emph{\small{The highlighted red block: the \hidden{} of the $2^{nd}$ token at the $2^{nd}$ layer.}}}
    \vspace{-0.2in}
    \label{fig:hcache:storage_sequence}
\end{figure}

\parabf{C2: Storage format.} Unlike \kvcache{} that owns dedicated GPU memory space,
hidden states are intermediate activations in temporary buffers repeatedly reused among transformer layers. 
Therefore, these \hidden{} should be dumped to host storage immediately once they are generated.
However, hidden states' generation and restoration order are inconsistent, making it hard to design 
an efficient storage format to manage \hidden{}. As shown in Figure~\ref{fig:hcache:storage_sequence}a, the 
\hcache{} is generated autoregressively, exhibiting a \emph{layer-before-token} order. 
In state restoration, instead, the \hidden{} of multiple tokens are restored as a batch at each layer 
(i.e., \emph{token-before-layer}, Figure~\ref{fig:hcache:storage_sequence}b). 
As a result, a storage format optimized for state saving (i.e., placing a token's \hidden{} at different layers in a continuous place)
incurs random, small-sized IOs at restoration, or vice versa.

\section{\sysname{} Design}

\label{design::overview}

\begin{figure}
    \centering
    \includegraphics[width=\linewidth]{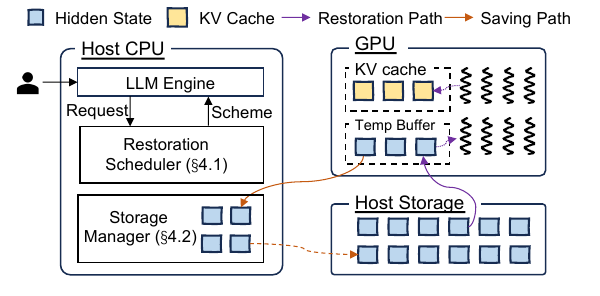}
    \vspace{-0.2in}
    \caption{\sysname{} Overview.}
    \vspace{-0.2in}
    \label{fig::design:overview}
\end{figure}

To address the above challenges, we further enhance \hcache{} with bubble-free restoration and efficient storage management. 
Figure~\ref{fig::design:overview} depicts the overall architecture of \hcache{} and 
how it interacts with an LLM inference engine.
When a user request arrives, the inference engine first decides whether the request's history states should be restored. 
If so, the inference engine utilizes the bubble-free restoration scheduler (\S\ref{design:subsec:bubble}) 
to generate an optimal restoration scheme 
to restore its \kvcache{} (to solve \textbf{C1}). With the restored \kvcache{}, 
the user request is further processed with prompt prefilling and token generation 
to generate an answer. During the prefilling and token generation phase, the storage manager (\S\ref{design:subsec:storage}) efficiently 
saves the newly computed \hidden{} to host storage (to solve \textbf{C2}). Since \sysname{} focuses on state restoration speed, we do not cache and reuse \kvcache{} in GPU.

By default, our storage manager utilizes SSDs as the backend storage devices. 
SSDs are cost-effective, offer substantial capacity, and deliver high I/O bandwidth, 
making them ideal for storing large volumes of \hidden{}. 
In environments lacking SSDs, host DRAM can be used as an alternative. 
Previous research has suggested using a hierarchical storage backend that combines host DRAM and SSDs~\cite{zuo2024as}. 
They also integrate prefetching and caching strategies, allowing frequently accessed contextual states to reside in the host DRAM. 
Contextual state caching is orthogonal to our work and can be incorporated to enhance performance further.

\subsection{Bubble-Free Restoration Scheduler}
\label{design:subsec:bubble}

The bubble-free restoration scheduler combines different restoration methods dynamically to eliminate 
pipeline bubbles. Specifically, the scheduler partitions a model's state across layers, with most states  
be managed via \hidden{}, while using a resource-complementary method (token recomputation or KV offload) 
for other states to fill in the bubble. 
For example, bubbles exist in transmission when the computation speed is slow, so we offload part 
of the model's state as \kvcache{} to prevent the need for computation.
\sysname{} determines an optimal partition scheme via offline profiling of the 
hardware characteristics. We further illustrate how a model's state should be partitioned (\S\ref{sec:design:state_partition}) 
and the detailed partition algorithm (\S\ref{sec:design:partition_algorithm}).

\subsubsection{State Partition Method}
\label{sec:design:state_partition}

When employing different restoration methods to eliminate pipeline bubbles, 
model states are partitioned like prior work~\cite{nara19pipedream,feng23mobius}, with each using a different restoration method. 
There are two approaches to partitioning the model state: token-wise partition and layer-wise partition. 

\begin{figure}
    \includegraphics[width=\linewidth]{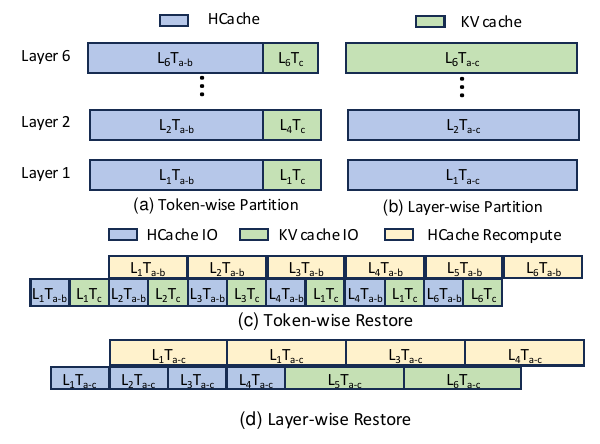}
    \vspace{-0.25in}
    \caption{Two Ways to Combine \sysname{} with \kvcache{}. \emph{\small{
    $\rm L_i$ and $\rm Tj$ indicate layer i and token j, respectively.
    }}}. 
    \vspace{-0.25in}
    \label{fig::design:hybrid-ways}
\end{figure}

Figure~\ref{fig::design:hybrid-ways}a shows a LLM model with 6 layers and a history context containing 3 tokens. 
With a token-wise partition scheme, these states are vertically split into two parts, 
where the first two tokens are managed via \hcache{} and the rest one is managed using a 
complementary method (e.g., KV offload). State restoration with token-wise 
partition is shown in Figure~\ref{fig::design:hybrid-ways}c. At layer $i$, KV recomputation from 
\hidden{} of the first two tokens is done concurrently with the transmission of layer $i+1$ (\hidden{} 
of the first two tokens and \kvcache{} of the 3rd token). In layer-wise partition, model states are horizontally 
partitioned. As shown in Figure~\ref{fig::design:hybrid-ways}b, the first 4 layers of an LLM are managed with 
\hcache{} while the KV cache of the last two layers is offloaded to the host. 
State restoration with token-wise partition is shown in Figure~\ref{fig::design:hybrid-ways}d, 
where the transmission of hidden states can occur continuously without requiring synchronization 
at each layer, provided it is faster than recomputation. Once complete, the \kvcache{} of the last two
layers is fetched into GPU memory.

\parabf{Performance considerations.}
In practice, token-wise partition does not always fully eliminate the bubbles in state restoration. 
We observe that the execution time of GEMM operations does not vary proportionally with the number of tokens involved.
The GEMM kernel in the cuBLAS library is well-optimized for standard matrix sizes. However, the token-wise 
partition can generate a scheme that forms an irregular matrix and is less optimized by cuBLAS.
As a result, executing a GEMM kernel with fewer tokens may consume a similar amount of time as one with more tokens. 
Even if we round the partition scheme to the nearest optimized size, bubbles still exist in the computation 
and transmission, resulting in suboptimal performance. 

Such a performance issue does not exist in layer-wise partition since all tokens at a layer are recomputed together. 
Furthermore, current LLM engines typically set a fixed mini-batch length to limit the intermediate buffer size and 
sequences larger than a mini-batch will be split and processed sequentially. 
Hence, We use the layer-wise partition and set the mini-batch length to be an optimized size in cuBLAS.

\subsubsection{State Partition Algorithm}
\label{sec:design:partition_algorithm}

We introduce the state partition algorithm to generate a bubble-free restoration scheme according to the hardware 
setups. We introduce two variables -- $L_H$ and $L_O$ -- to indicate the 
number of model layers managed with \hcache{} and other methods, respectively.

Specifically, on a platform with faster computation speed, \hcache{} is combined with token recomputation, 
where the first $L_O$ layers are restored with token recomputation, and the 
remaining $L_H$ layers with \hcache{}. When recompute of the first $L_O$ layers. The \hidden{} of the 
latter layers are prefetched. When token recomputation finishes, \hcache{} restoration is started to continue the computation. 
On a platform with faster I/O transmission speed, \hcache{} is combined with KV offload. 
For the first $L_H$ layers, we copy their \hidden{} and recompute them to restore the \kvcache{}. 
Since the  I/O transmission of \hidden{} is faster, it will finish earlier than the computation, 
so we copy the \kvcache{} of the rest $L_O$ layers in the remaining time. 

Our state partition algorithm sets $L_H$ and $L_O$ such that data transmission and computation 
finish almost simultaneously, thus eliminating bubbles. To solve $L_H$ and $L_O$, we profile the transmission and computation speed of a 
specific hardware platform offline, where the transmission time of \hidden{} is $IO_H$, 
transmission time of \kvcache{} is $IO_{KV}$, computation time from token is  $C_{Token}$, from \hidden{} to \kvcache{} is $C_H$.
We formulate the bubble-free target as a min-max optimization problem. Below is an example of combining \hcache{} with \kvcache{}:
\begin{equation*}
\begin{aligned}
    \argmin_{L_H,L_O} \max\ &(C_H*L_H,IO_H*L_H+IO_{KV}*L_O) \\
    \text{ subject to}&\ L_H+L_O = N_{Layer}
\end{aligned}
\end{equation*}
Given a hardware configuration, $L_H$ and $L_O$ are derived via:
\begin{equation*}
\begin{aligned}
L_H &= \lceil \frac{N_{Layer}* IO_{KV}}{IO_{KV}+C_H-IO_H} \rceil, ~~~\mathit{if}\  C_H > IO_H \\
L_H &= \lceil \frac{N_{Layer}* C_{Token}}{C_{Token}+IO_H-C_H} \rceil, ~~~\mathit{if}\  C_H\le IO_H \\
L_O &= N_{layer}-L_H
\end{aligned}
\end{equation*}

\subsection{\sysname{} Storage Manager}
\label{design:subsec:storage}
The storage manager is tasked with managing contextual states in host storage. 
We have adopted a storage format optimized for state restoration, as reducing TTFT 
is our primary design goal. Additionally, we introduce a two-stage state-saving mechanism 
to mitigate the overhead associated with saving newly generated states.
Note that with our state partition algorithm in \S\ref{sec:design:partition_algorithm},
the contextual states at each model layer are stored either in \hidden{}, \kvcache{}, or original tokens.
Here, we focus on the management of \hidden{} since the \kvcache{} can be managed similarly.

\subsubsection{Storage Format}
\label{sec:design:storage:format}
We introduce a storage format designed for fast restoration. 
Since we adopt a layer-wise approach to restore the states for the history context, 
the \hidden{} of all tokens from the same layer are copied together. 
Conceptually, the \hidden{} of tokens in the same layer is colocated in a continuous large data block to maximize transferring bandwidth. In practice, we split the tokens of one layer into multiple fix-sized (64 tokens) chunks. The chunks of a layer are distributed 
on multiple SSDs following the round-robin manner. The chunk-based storage format is designed for the following reasons:
First, when multiple SSDs are present in the system, we should distribute the 
data of a layer across all SSDs instead of a large continuous block on a single SSD. This can facilitate parallel I/O operations and 
enhance the transmission speed of one layer through bandwidth aggregation. 
Second, LLM generates tokens autoregressively, whose output length is unpredictable~\cite{yu2022orca,stoica2023vllm}. 
This forbids us from reserving the storage space for a layer's \hidden{} as an entire buffer -- 
reserving space according to its maximum length would result in serious internal fragmentation. 
\subsubsection{Two-Stage State Saving}
\label{sec:design:storage:aync}
 
Next, we explore how the storage manager saves newly generated hidden states using a chunk-based storage format.
As we mentioned before, hidden states are intermediate activations generated at each layer, 
if they are not saved in a timely manner, computation tasks will be blocked.
In the token generation phase with continuous batching enabled, a batch generates \hidden{} of different sequences, 
which should be organized into separate chunks before being saved. 
Directly transferring these hidden states to different chunks at host storage can lead 
to numerous small write operations, potentially compromising the efficiency of the generation task.

We adopt a two-stage strategy to address this problem. 
First, the \hidden{} of the batch, encompassing tokens from various sequences, are collectively copied to the 
host memory using a single \texttt{cudaMemcpy} call. This method effectively snapshots the 
hidden states to the host, allowing the GPU memory buffer to be properly reused for subsequent layers. 
Second, a host daemon running on the CPU orchestrates chunk management. It copies \hcache{} to appropriate chunk buffers 
and flushes them to storage devices.

In this way, the overhead of saving \hidden{} to host storage is moved off the 
critical path of LLM processing; meanwhile, small-sized writes are reorganized into large chunks,  
which host storage (e.g., SSDs) favors.

\section{Implementation}
We implement \sysname{} with 5731 lines of code in Cuda, C++, and Python. Our implementation is based on 
DeepSpeed-MII v0.2.0~\cite{MII}, a state-of-the-art LLM serving system that supports the 
SplitFuse~\cite{agrawal2024taming} scheduling algorithm. 
We add support restoring from \hidden{} to \kvcache{} in DeepSpeed-MII. 
Specifically, we use the cuBLAS library to project the \hidden{} to primeval KV values. 
Following prior work~\cite{zuo2024as}, we write a custom kernel to apply the 
ROPE position embedding~\cite{su2024roformer} to the recomputed KV values and copy the result to the \kvcache{}.

\parabf{Request scheduling.}
The scheduler of \sysname{} derives from the continuous batching~\cite{deepspeed2024fastgen} method. There are two main phases in continuous batching, namely prefilling and decoding. \sysname{} adds an extra restoration phase. The request arrived in the system with evited history \kvcache{} will first execute this phase with bubble-free restoration. When the restoration phase is finished, The prefill phase is then conducted using the request's newly arrived prompt. Finally, the request is added to the decoding batch. We also apply the SplitFuse~\cite{agrawal2024taming} scheduling algorithm to opportunistically fuse prefill with decoding generation to mitigate the interference between prefilling and decoding.

\parabf{Pipelined state transmission.}
On the GPU side, we use dedicated streams for state transmission 
-- one for the upstream state transmission and the other for the downstream state snapshot. 
We use cudaEvent to coordinate the ordering of operations among different streams. 
For example, the snapshot of one layer's \hidden{} is followed by an event, which is 
waited by the first compute operator of the next layer so as to keep the buffer a 
consistent version when accessed.

\parabf{Storage management.} To save \hidden{}, we use 8 background threads on the host side 
to collect dumped \hidden{} and append them to corresponding data chunks.
Once a chunk is fully populated, it is promptly written to the NVMe device.
To restore \hidden{}, similar to FlashNeuron~\cite{lee2021flashneuron}, 
we co-use SPDK~\cite{yang2017spdk} and GDRCopy~\cite{gdrcopy} to realize GPUDirect Storage, 
which transfers data between the GPU and SSD directly. 
Specifically, we pin several memory buffers on the GPU's BAR (Base Address Register) space in advance;
when the GPU reads data from the SSD, we set the NVMe read command's destination address as the exposed GPU BAR address, 
thus achieving a direct P2P copy, which eliminates redundant memory copies with minimized I/O latencies.

\parabf{Multi-GPU support.} We also add support for \sysname{} to 
serve large models that run on multiple GPUs. With tensor parallelism, 
each GPU node should have a full version of \hidden{} to compute 
the \kvcache{}. To avoid redundant read of the \hidden{} from different GPUs, we 
let all GPUs read \textit{disjoint} shards (split by token) 
of hidden states concurrently. 
This operation can aggregate the read bandwidth of all GPUs with no read amplification. 
Then, GPUs run an all-gather communication operator to reconstruct the full hidden state from shards. 
Since the GPU is connected via the fast NVlink interconnect, the all-gather operation will only incur a small overhead compared with the transmission part. With pipeline parallelism. The restoration of each layer's \hcache{} is independent of each other. 
So, each GPU node can fetch the \hidden{} of its responsible layers concurrently 
and recompute it to attain the \kvcache{} of corresponding layers.

\begin{figure*}[t!]
    \centering
    \includegraphics[width=\linewidth]{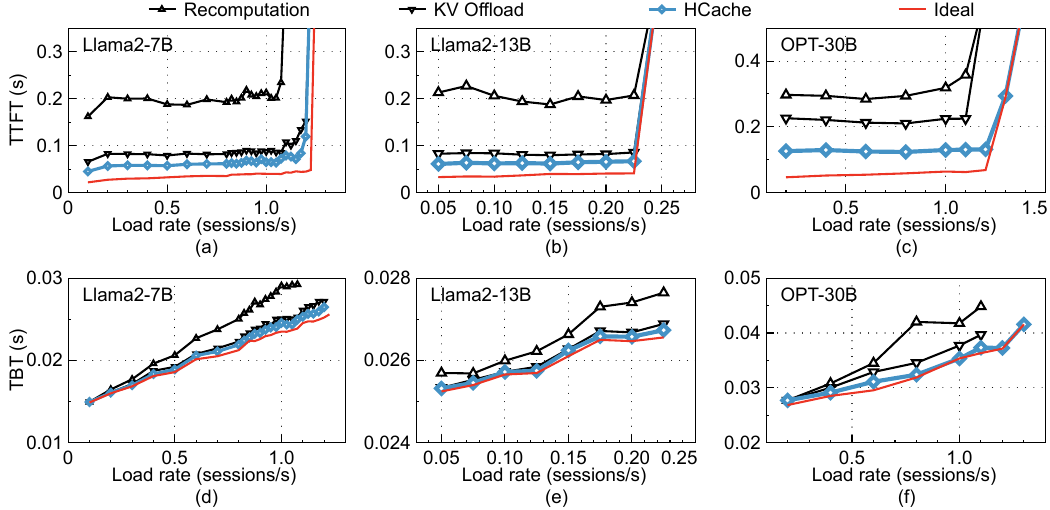}
    \vspace{-0.15in}
    \caption{Overall Performance Results On ShareGPT4 Trace.}
    \label{fig::expr::sharegpt}
\end{figure*}
\section{Evaluation}

We evaluate \sysname{} to answer the following questions:
\begin{compactitem}[\scriptsize{$\bullet$}]
\item How does \sysname{} perform in terms of TTFT and TBT compared to state-of-the-art state restoration methods?
\item How does \sysname{} perform on platforms with various GPUs, storage devices, and context length?
\item How do \sysname{}'s individual techniques affect \hcache{}'s performance?

\end{compactitem}

\parabf{Testbed.} 
Unless otherwise stated, all our experiments are conducted on a server equipped with 
4$\times$ A100-40G SXM4 connected via NVLink. The host has 2$\times$ AMD EPYC 7642 
CPUs, 256G DDR4 memory, and $4\times$ Samsung PM9A3 4TB enterprise SSDs. In our sensitivity experiments, 
we change the hardware configuration by using cloud servers equipped with different GPUs.
For these cloud servers, we directly use the host DRAM as the storage backend. 
The hardware characteristics of these GPUs are shown in Table~\ref{tab:eval: hardware}.

\begin{table}[b]
    \centering
    \setlength{\tabcolsep}{2.2mm}{
        \begin{tabular}{cccccc}
            \arrayrulecolor{black}\hline
            \arrayrulecolor{black}\hline
            \textbf{GPU} &\textbf{HBM Size} & \textbf{FLOPS$^{\star}$} & \textbf{Transmission Speed} \\ 
            \hline
            A100 & 40G & 312T & 32GB/s\\ 
            A30 & 24G & 165T & 32GB/s \\ 
            4090 & 24G & 330T & 32GB/s  \\
            L20 & 48G & 120T & 32GB/s  \\
            H800 & 80G & 990T & 64GB/s  \\
            \arrayrulecolor{black}\hline
        \end{tabular}
    }
    \caption{Hardware Characteristics of Different Platforms. \emph{\small{
    Note that $^{\star}$ indicates the FLOPS of FP16 operations.
    }}}
    \vspace{-0.15in}
    \label{tab:eval: hardware}
\end{table}

\parabf{Models}:  We use Llama2-7B, Llama2-13B, OPT-30B as our test models. We 
expand the maximum context length of these models to 16K to accommodate L-Eval benchmarks and long conversation history.
For Llama2-7B/13B, we use a single A100 GPU to serve them. For OPT-30B, we run it 
on 4 A100 GPUs with tensor parallelism unless otherwise stated. 
We use the ShareGPT4 and L-Eval described in \S\ref{sec:bg:stateful} as our test traces to imitate the real-world LLM use cases like multi-round conversation chatbot or RAG application. 

\parabf{Metrics.} When evaluating the overall performance, we report TTFT (Time to First Token), 
which is the duration of the restoration and prefill phase, and TBT (Time between Token), which 
represents the average time taken to generate a token for each request except for the first token. TTFT 
is highly related to user experiences, and we mainly report TTFT to show \sysname{}'s performance improvements.
We report TBT to prove that \sysname{} poses negligible impact on decoding requests. 
In other case studies, we run restoration requests with a batch size equal to 1, so we report the restoration speed instead, 
which is defined as the number of restored historical tokens divided by the restoration time. 

\parabf{Baselines.} We compare \sysname{} to the following baselines.
\underline{Recomputation}: DeepSpeed-MII\cite{MII} is an LLM serving system that supports the SplitFuse scheduling 
algorithm and uses PagedAttention to manage \kvcache{}. We use it as the baseline for token recomputation.
\underline{KV offload}: AttentionStore~\cite{zuo2024as} offloads \kvcache{} to multi-tiered secondary storage. We implement it on top of DeepSpeed-MII to serve as the KV offload baseline. We do not implement the decoupled position embedding because we already expand the context length for models.
\underline{Ideal}: We also implement an ideal system without the restoration overhead.
Here, we do not keep all \kvcache{}s in GPU memory; instead, we allocate the placeholder KV values in GPU, which are used repeatedly across all user requests.
This achieves the theoretical lower bounds for the TTFT and TBT metrics, assuming all the history KV cache is cached on GPU and reused.

\subsection{Overall Performance }

\subsubsection{Multi-Round Conversation}
We evaluate the overall performance of \sysname{} against baseline methods 
in the multi-round conversation setting with ShareGPT4. 
In this experiment, we use 4 SSDs as the storage backend. 
Following prior work~\cite{zuo2024as,stoica2023vllm}, we set the arrival time of 
different sessions with the Poisson distribution. 
The interval between conversation rounds in one session is set to 30s. All restoration methods do not reuse the KV cache on GPU for a fair comparison. The \kvcache{} are evicted when one round of conversation ends.

\parabf{TTFT and throughput}.
Figures~\ref{fig::expr::sharegpt}a-c show the TTFT under different request rates in the multi-round conversation. 
Overall, \sysname{} can provide 1.27-1.90$\times$ TTFT speedup compared with the KV offload and 2.21-3.57$\times$ better than token recomputation.
For 7B/13B models, the one running GPU has 4 SSDs, providing sufficient IO bandwidth for state restoration. Hence, the KV offload method is fast, but our methods can still be 1.27-1.60$\times$ faster than it.
Instead, OPT-30B runs on 4 GPUs, each with one SSD, so 
the restoration time of OPT-30B is longer, and \sysname{} can achieve up to 1.90$\times$ TTFT speedup compared with KV offload.

The 13B model has a lower peak throughput compared to 7B/30B models. 
This is bounded by the limited GPU memory for the \kvcache{}. 
The throughput of different restoration methods is nearly the same in this case.
For the 7B/30B model, \sysname{} can sustain up to 11\% more requests than 
KV offloading since the restoration cost for \sysname{} is lower.

\parabf{TBT}.
Figures~\ref{fig::expr::sharegpt}d-f show the TBT of different models in the multi-round conversation setting. 
Compared with the ideal case, in which no restoration is needed, \sysname{}'s TBT is at most 4\% higher, 
almost the same as the ideal case. The low TBT overhead benefits from the high restoration speed of \sysname{} 
, which reduces GPU time spent on the restoration, so the token generation is stalled long. 
\begin{figure}
    \centering
    \includegraphics[width=\linewidth]{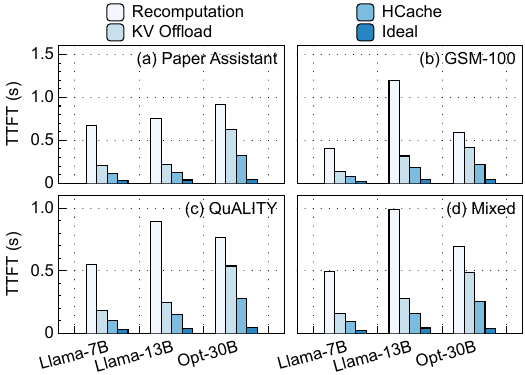}
    \vspace{-0.25in}
    \caption{Performance of Long-Context Applications.}
    \vspace{-0.2in}
    \label{fig::expr::l-eval}
\end{figure}

\subsubsection{Long-Context Applications}
We evaluate the overall performance of \sysname{} against baseline methods in the 
long-context applications setting with the L-Eval trace. Since the GPU HBM can 
only serve 1-3 long-context requests, we evaluate different models with a batch 
size equal to 1. Figure~\ref{fig::expr::l-eval} shows the TTFT of three representative 
subtasks and the sampled 200 requests from the trace (i.e., mixed). \sysname{} can 
achieve 1.62-1.93$\times$ speed up for TTFT compared over KV offload and 2.66-5.73$\times$ over token recomputation. 
It is worth noticing that these tasks' history length spans within a large range from 4K to 16K, demonstrating \sysname{}'s 
excellent scalability.

\subsubsection{In-Depth Analysis}
In Table~\ref{tab:eval:storage}, we report the schedule results of \hcache{}, in terms of how 
model layers are managed and the per-token storage space consumption. We also 
report the storage cost of KV offload as a comparison.  
Specifically, the 7B model on one A100 has balanced speed for recomputation from 
\hidden{} into \kvcache{} and the transmission of \hidden{}, so we use \hidden{} 
for 31 layers and only transmit the \kvcache{} for one layer to fill in the bubble. 
For the 13B and 30B model, \sysname{} use \hidden{} for more than 80\% layers, 
while the rest of the layers are managed via different resource-complementary methods. 
Regarding space consumption, the size of \hidden{} for one token is half that of \kvcache{}.
Combined with the zero-bubble scheduler, some layers may not even need to be stored because they can be 
recomputed from tokens. As a result, \sysname{}'s per token storage space is 1.92-2.40$\times$ lower than KV offload. To achieve a balanced speed between computation and transmission using only hidden states, approximately 24GB/s, 21GB/s, and 37GB/s of storage bandwidth are needed for the 7B, 13B, and 30B models, respectively.

\begin{table}[t!]
    \centering
    \setlength{\tabcolsep}{3.2mm}{
        \begin{tabular}{c|c|c|c}
            \arrayrulecolor{black}\hline
            \arrayrulecolor{black}\hline
            \multirow{2}{*}{\textbf{Model}} & \multirow{2}{*}{\textbf{Schedule}} & 
            
            \multicolumn{2}{c}{\textbf{Per Token Storage Cost}} \\
            \cline{3-4}
             &  &\textbf{\sysname{}} & \textbf{KV Offload}\\
            \hline
            7B & 31 H + 1 KV  & 132 KiB & 256 KiB\\ 
            13B & 36 H + 4 KV & 210 KiB & 400 KiB\\ 
            30B & 40 H + 8 RE & 280 KiB & 672 KiB \\
            \arrayrulecolor{black}\hline
        \end{tabular}
    }
    \caption{Scheduling Results and Storage Cost.\emph{\small{
    H, KV, and RE indicate \hidden{}, \kvcache{}, and recomputation, respectively.
    }}}
    \vspace{-0.25in}
    \label{tab:eval:storage}
\end{table}

\subsection{Sensitivity Analysis}
\label{sec:eval:sensi}
The state restoration speed of an LLM is highly relevant to the model size, IO transmission bandwidth, 
computation power, and historical context length. We vary them to analyze the sensitivity of \hcache{} against baseline methods.

\begin{figure*}[t!]
    \centering
    \includegraphics[width=0.95\linewidth]{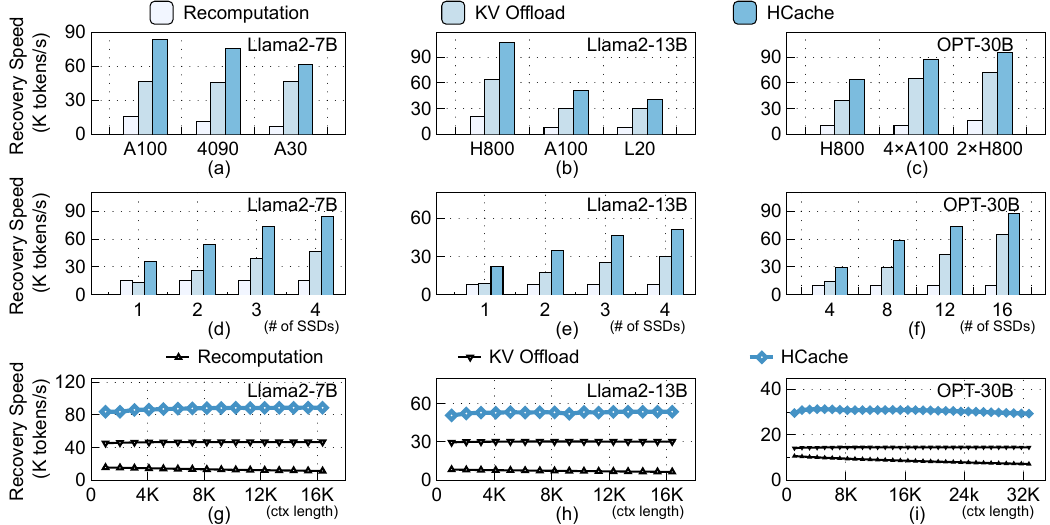}
    \vspace{-0.05in}
     \caption{Sensitivity Analysis by Various Factors. \emph{\small{
    (a-c): Varying GPU computation capability and using DRAM as the storage backend.
    (d-f): Varying number of SSDs in the default testbed.
    (g-i): Varying context length with the default testbed.
    }}}
    \vspace{-0.05in}
    \label{fig::expr::sensitivity}
\end{figure*}
\subsubsection{Varying GPU Devices}

The restoration phase of \sysname{} includes the projection of \hidden{} to \kvcache{}, so its performance 
is sensitive to the computation power of GPU. We evaluate the restoration speed of \sysname{} and baselines on various GPU platforms. We use host DRAM as the storage backend so that the transmission speed is not the bottleneck. Figures~\ref{fig::expr::sensitivity}a-c shows that \sysname{} outperforms KV offload by 1.33–1.81× and Recomputation by 5.04–9.05× in restoration speed across different platforms. 
Platforms with low computation capability are unfriendly to \sysname{}, such as running a 7B model on an A30 GPU. 
In this case, \hcache{} computation is longer than transmission and makes the speed up less than 2$\times$. 
Even so, \sysname{} still restores 1.33$\times$ faster than KV offload. 
On platforms with faster computation speeds, \sysname{} shows improved performance, achieving a 1.66$\times$ increase on A100 GPU 
and a 1.77$\times$ increase on the H800 GPU for 13B models compared to KV offload.
Compared to token recomputation, \sysname{} achieves 5.04-9.05$\times$ speed up on different GPUs. This is attributed to the 6$\times$ theoretical speedup against token recomputation by skipping the attention and FFN modules.

\subsubsection{Varying Storage Bandwidth}

LLM state restoration is sensitive to the transmission speed. We evaluate the restoration speed of \sysname{} and baselines with different numbers of disks. One PM9A3 SSD provides a read bandwidth of 6.9 GB/s, and using 4 disks can saturate the upstream PCIe bandwidth of the A100 GPU. For the 30B model, we use DRAM to simulate 8-16 disks since 
our platform does not have enough PCIe slots. The history length is set to 1024.

Figures~\ref{fig::expr::sensitivity}d-f present the sensitivity evaluation results related to transmission speed. 
In terms of the restoration speed across various disk configurations,
\sysname{} significantly outperforms KV offload, delivering a 1.7 to 2.6$\times$ improvement 
and surpasses recomputation with a 2.3-6.1$\times$ enhancement. 
For platforms with fewer disks, IO transmission can be slower than computation tasks; 
based on our theoretical analysis, \sysname{} can perform at least $2\times$ better, and 
the bubble-free scheduler can bring extra speed up. For example, on the platform with one SSD per GPU,
the overall improvement of \sysname{} over KV offload is 2.09-2.66$\times$.
For platforms with more disks, instead, the recomputation operation in \hcache{} can be longer than the transmission in this setting, so the speed up can be less than 2$\times$. However, the recomputation overhead is much lower than 
token recomputation. The overall speedup is 1.33-1.81$\times$ on 7B-30B models compared with KV offload and 5.46-6.08$\times$ compared with recomputation from the token.

\subsubsection{Varying Sequence Length}
We further investigate the impact of sequence length on \sysname{}'s restoration speed. 
Following the same configuration as the overall experiment, we test different models' restoration speeds on A100 GPUs 
with 4 SSDs by varying the history length. 
As shown in Figures~\ref{fig::expr::sensitivity}g-i, token recomputation can not scale well with the increasing history length; the restoration speed of a 7B model drops by 28\% as the history length increases from 1K to 16K. 
Token recomputation's quadratic complexity attention mechanism incurs a high overhead in contexts with a long history. 
The \kvcache{} can scale well with different history lengths because its transmission size is proportional to the number of tokens. \sysname{} also scales well with different history lengths for two reasons. First, the transmission size of \hidden{} and recomputation cost from it to \kvcache{} are all proportional to the number of tokens (see ~\S\ref{sec:hcache:compare}). Second, \sysname{} has the bubble-free restoration scheduler and may opportunistically combine the usage of the token recomputation to fill in the bubble in the short context. With long historical contexts, the scheduler detects that token computation is very expensive and falls back to the \hcache{}-only approach. 

\begin{figure}
    \centering
    \includegraphics[width=\linewidth]{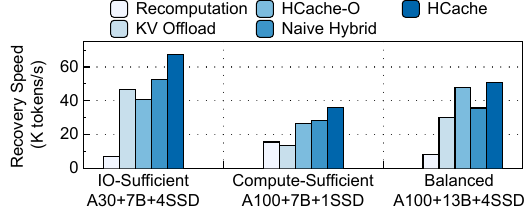}
    \caption{Ablation Study on Bubble-Free Scheduler.}
    \label{fig::expr::ablation-opt}
\end{figure}

\subsection{Ablation Study}

\subsubsection{Bubble-Free Scheduler}
\label{sec:eval:hybrid}
We first compare different scheduling methods for state restoration. 
We implement two variants of \sysname{}, including \sysname{}-O and Naive hybrid. 
\sysname{}-O only uses \hidden{} for state restoration without the bubble-free scheduler. 
Naive Hybrid uses the bubble-free restoration scheduler to mix token recomputation and KV offload without using \hidden{}. 
We compare them under three hardware settings: balanced, compute-, and IO-sufficient.

The results are shown in Figure~\ref{fig::expr::ablation-opt}. 
Without \hidden{}, Naive hybrid is the best method that uses both compute and transmission resources. 
\sysname{} outperforms this approach by 1.28-1.42$\times$. Both methods have no bubbles, 
so the performance gain of \sysname{} comes from the \hidden{} as it has theoretically lower computational and IO overhead.
 
The bubble-free scheduler plays a key role on resource-skewed platforms. 
\hcache{}-O incurs pipeline bubbles and waste resources. 
For example, in the IO-sufficient setting, the bubble in the \hcache{}-O lets it be 13\% slower 
than KV offload because the recomputation in \sysname{} is slow and IO resources are wasted.  
The bubble-free scheduler can integrate \hcache{}-O with resource-complementary methods, 
thus improving the speed of \hcache{}-O by 1.35-1.64$\times$ in skewed hardware configuration 
and making \sysname{} consistently perform better than KV offload by 1.45-2.66$\times$.

\begin{figure}
    \centering
    \includegraphics[width=\linewidth]{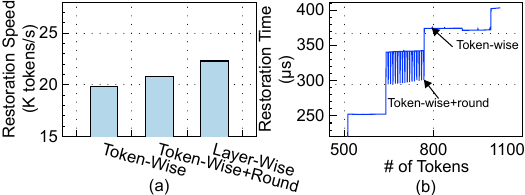}
    \vspace{-0.25in}
    \caption{Ablation Study on State Partition Methods.}
    \vspace{-0.2in}
    \label{fig::expr::ablation-layout}
\end{figure}
\subsubsection{State Partition Methods}
We compare different methods to partition the state of the model. We measure the restoration speed 
of the 13B model using history contexts with 1024 tokens. We run the model on one A100 with one SSD. 
With the layer-wise partition, our algorithm produces a scheme that uses \hidden{} for 31 layers 
and token recomputation for the remaining 9 layers. Instead, a naive token-wise partition uses \hidden{}
for 794 tokens and token recomputation for the remaining 230 tokens. A round-up optimization over the token-wise 
partition uses the nearest optimal size (i.e., 768) to manage tokens with \hcache{}. 
As Figure~\ref{fig::expr::ablation-layout} shows, the restoration speed with the naive token-wise partition is 
12\% slower. While the round-up optimization can improve performance by issuing a more performant cuBLAS kernel, 
it is still 7\% slower than layer-wise partition because of the unbalanced workload in computation and transmission. 
As auxiliary information, we also report the restoration time of GEMM operation in one layer with varying numbers of 
tokens (see Figure~\ref{fig::expr::ablation-layout}b). 
\begin{figure}[t!]
    \centering
    \includegraphics[width=\linewidth]{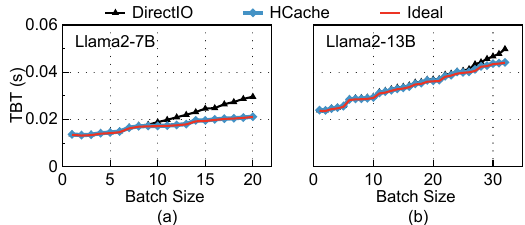}
    \vspace{-0.25in}
    \caption{Ablation Study on Two-Stage \hcache{} Saving.}
    \vspace{-0.3in}
    \label{fig::expr::ablation-saving}
\end{figure}
\subsubsection{Two-stage State Saving}
To evaluate the effectiveness of the two-stage \hidden{} saving strategy. We implement a variant of \sysname{}, which directly saves \hidden{} to SSDs (i.e., DirectIO in Figure~\ref{fig::expr::ablation-saving}). 
We test the TBT with different numbers of sequences in the decoding batch and set the history length of each to 512.  

As shown in the figure, \sysname{}'s TBT is consistent with the ideal case. This is because \texttt{cudaMemcpy} can efficiently copy the \hidden{} to host DRAM. DirectIO achieves a similar effect when the batch size is small because the I/O time is less than one layer's decoding time. However, when the batch size is large, DirectIO can not store one batch's \hidden{} to the device on time and will stall the decoding phase. DirectIO's TBT can be 34\% higher with the 7B model when the batch size reaches 16. With the 13B model, 
DirectIO's TBT is less significant because each layer's decoding time becomes longer. 
However, as the batch size increases to 32, DirectIO still increases TBT by 13\%. 

It is important to note that no stalling occurred during hidden state snapshots in our experiments. For example, prefilling 1,024 tokens in a single layer of the Llama2-13B model generates approximately 10MB of hidden states in 3ms. This results in an equivalent bandwidth of 3GB/s, which is lower than the PCIe bandwidth. The hidden state generation speed during the decoding phase is even slower.

\subsection{Performace with on GPU KV Reusing}
Real-world LLM serving systems \cite{zheng2023efficiently,gim2024prompt} typically reuse the \kvcache{} on GPU. We evaluate the TTFT of \sysname{} against baseline state restoration methods, along with a  GPU-resident KV cache.

In this experiment, we use the L-Eval dataset and reuse the \kvcache{} for long-context tasks on the GPU, employing an LRU-based cache, as done in prior work \cite{zheng2023efficiently}. When a cache miss occurs on the GPU, the \kvcache{} is restored using different state restoration methods. Following prior work~\cite{song2023ugcache,xie2022fleche,xie2023petps}, we synthetic the arrival pattern of contexts in the L-Eval dataset with varying Zipfian skewness values ($\alpha$). We evaluate the 7B model with 4 SSDs, and we report both the cache hit ratio and TTFT under different skewness levels.

\begin{figure}
    \centering
    \includegraphics[width=\linewidth]{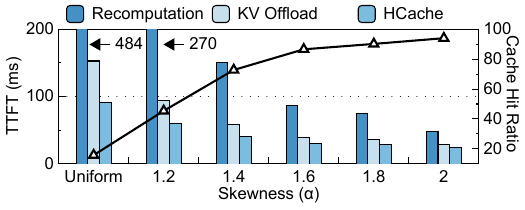}
    \vspace{-0.25in}
    \caption{Performace with on GPU KV Reusing}
    \vspace{-0.2in}
    \label{fig::expr::cache}
\end{figure}

The results are shown in Figure~\ref{fig::expr::cache}. With a uniform arrival pattern, the cache hit ratio is 15\%, and the TTFT is similar to the overall experiment, which does not incorporate caching. In the uniform setting, \sysname{} is 1.67$\times$ faster than KV offloading. As the skewness increases from uniform to $\alpha = 2.0$, the cache hit ratio rises to 94\%, allowing the GPU cache to significantly reduce TTFT by 3.76$\times$ to 10.03$\times$. A high cache hit ratio diminishes the performance gains of \sysname{} due to fewer state restoration events. Nevertheless, \sysname{} remains 1.15$\times$ faster than KV offloading and 1.98$\times$ faster than recomputation in high-skewness workloads.

\section{Related Work}
\parabf{Stateful LLM optimizations.}
Recent work optimized for stateful LLMs all maintain the \kvcache{} as is. PromptCache 
~\cite{gim2024prompt}, SGLang~\cite{zheng2023efficiently} and 
ChunkAttention~\cite{ye2024chunkattention} proposes to cache and reuse the on-chip \kvcache{} 
for shared tokens across different requests. They optimize the GPU cache hit path and are complementary to \sysname{} which optimize the cache miss path.  For larger than GPU memory \kvcache{} management, 
AttenionStore~\cite{zuo2024as} offloads the \kvcache{} to tiered host storage, including DRAM and SSDs. 
RAGCache\cite{jin2024ragcache}, Pensieve\cite{yu2023stateful}, and 
APIServe~\cite{yiying2024apiserve}  maintain a multi-tier \kvcache{} storage 
system across GPU memory and host DRAM based on hotness and restoration overhead. 
FlexGen~\cite{sheng2024flexgen} is an offline inference system that offloads both 
\kvcache{} and output activations of each layer to facilitate batched inference. 
Our work is distinguished from them by exploiting \hidden{}, and the above 
caching mechanisms can be further applied in our system to boost performance.

\parabf{LLM state compression.}
One alternative to reducing LLM's state restoration time is to compress the \kvcache{}, but at the cost of 
compromised model accuracy. Instead, \hcache{} is a \emph{lossless} method for state restoration.

\noindent
\underline{Quantization}~\cite{gray1998quantization,liu2024kivi,dong2024qaq,hooper2024kvquant,zhang2024kv,
yang2024no,kang2024gear} is the representative methods for \kvcache{} compression. 
These methods compress the \kvcache{} from 16-bit into 4-bit or even lower. 
CacheGen~\cite{liu2024cachegen} is the most relevant work that combines KV offload with quantization. 
CacheGen mitigates the KV transferring overhead by reducing its size with quantization-based compression algorithms. 
These quantization-based methods can be applied in \hcache{} to reduce the size of \hidden{}. 
Moreover, LLM weight quantization methods~\cite{zhao2023atom,frantar2022optq,pmlr-v202-xiao23c,lin2023awq} 
are also applicable to reduce computing overhead and can be combined 
with \hcache{} to improve state restoration speed.

\noindent
\underline{Token drop methods}. Some recent work~\cite{zhang2024h2o,mu2023learning} drops 
the \kvcache{} of less important tokens directly 
based on attention scores. These methods reduce the \kvcache{} size by reducing the 
number of tokens that need to be stored. 
They can also be integrated into \hcache{} to reduce the transmission and computation overhead.

\noindent
\underline{MQA and GQA}. Algorithm improvements like MQA~\cite{shazeer2019fast} and GQA~\cite{ainslie2023gqa} have been designed to reduce the sizes of \kvcache{} by using low-rank KV representation. These methods can apply to \hcache{} by first projecting the \hidden{} into a low-rank representation and then storing it. This involves changing the model structure, which is beyond the scope of this paper. \hcache{} can currently support LLMs using the MHA mechanisms without any change of the model structure, including 20,000 famous LLM models on Hugging Face like Llama2~\cite{touvron2023llama}, Gemma~\cite{gemmateam2024gemma}, Phi2~\cite{javaheripi2023phi}, and Qwen1.5. 
\section{Conclusion}
We present \sysname{}, a novel method for LLM state restoration. \sysname{} stores the intermediate activation of LLM and combines the transmission and computation resources to restore the LLM state with low overhead.  
To further improve \hcache{}'s performance, we propose bubble-free restoration to use resource-complementary 
methods to improve the restoration speed, and introduce a chunk-based storage layout for \hcache{} and save the \hcache{} using a two-stage strategy. We evaluate \sysname{} with a range of models and hardware platforms. The experimental results show that \sysname{} can reduce TTFT  by up to 1.93$\times$ compared with KV offload and 5.73$\times$ compared with token recomputation. Moreover, \sysname{} can also reduce the storage cost by 1.92-2.4$\times$.

\section*{Acknowledgements}
We sincerely thank our shepherd Rong Chen and anonymous reviewers for their feedback and suggestions. This work is supported by the National Key R\&D Program of China (Grant No. 2022YFB4500302) and the National Natural Science Foundation of China (Grant No. U22B2023, 62202255).

\newpage

\bibliographystyle{ACM-Reference-Format}
\bibliography{ref}


\begin{thebibliography}{67}


\ifx \showCODEN    \undefined \def \showCODEN     #1{\unskip}     \fi
\ifx \showDOI      \undefined \def \showDOI       #1{#1}\fi
\ifx \showISBNx    \undefined \def \showISBNx     #1{\unskip}     \fi
\ifx \showISBNxiii \undefined \def \showISBNxiii  #1{\unskip}     \fi
\ifx \showISSN     \undefined \def \showISSN      #1{\unskip}     \fi
\ifx \showLCCN     \undefined \def \showLCCN      #1{\unskip}     \fi
\ifx \shownote     \undefined \def \shownote      #1{#1}          \fi
\ifx \showarticletitle \undefined \def \showarticletitle #1{#1}   \fi
\ifx \showURL      \undefined \def \showURL       {\relax}        \fi
\providecommand\bibfield[2]{#2}
\providecommand\bibinfo[2]{#2}
\providecommand\natexlab[1]{#1}
\providecommand\showeprint[2][]{arXiv:#2}

\bibitem[sha(2024)]%
        {sharegpt4}
 \bibinfo{year}{2024}\natexlab{}.
\newblock \bibinfo{title}{Sharegpt4 Dataset}.
\newblock \bibinfo{howpublished}{\url{https://huggingface.co/datasets/openchat/openchat_sharegpt4_dataset}}.
\newblock
\newblock
\shownote{[Computer software]}.


\bibitem[Abhyankar et~al\mbox{.}(2024)]%
        {yiying2024apiserve}
\bibfield{author}{\bibinfo{person}{Reyna Abhyankar}, \bibinfo{person}{Zijian He}, \bibinfo{person}{Vikranth Srivatsa}, \bibinfo{person}{Hao Zhang}, {and} \bibinfo{person}{Yiying Zhang}.} \bibinfo{year}{2024}\natexlab{}.
\newblock \bibinfo{title}{APIServe: Efficient API Support for Large-Language Model Inferencing}.
\newblock
\newblock
\showeprint[arxiv]{2402.01869}~[cs.LG]


\bibitem[Agrawal et~al\mbox{.}(2024)]%
        {agrawal2024taming}
\bibfield{author}{\bibinfo{person}{Amey Agrawal}, \bibinfo{person}{Nitin Kedia}, \bibinfo{person}{Ashish Panwar}, \bibinfo{person}{Jayashree Mohan}, \bibinfo{person}{Nipun Kwatra}, \bibinfo{person}{Bhargav Gulavani}, \bibinfo{person}{Alexey Tumanov}, {and} \bibinfo{person}{Ramachandran Ramjee}.} \bibinfo{year}{2024}\natexlab{}.
\newblock \showarticletitle{Taming {Throughput-Latency} Tradeoff in {LLM} Inference with {Sarathi-Serve}}. In \bibinfo{booktitle}{\emph{18th USENIX Symposium on Operating Systems Design and Implementation (OSDI 24)}}. \bibinfo{publisher}{USENIX Association}, \bibinfo{address}{Santa Clara, CA}, \bibinfo{pages}{117--134}.
\newblock
\showISBNx{978-1-939133-40-3}
\urldef\tempurl%
\url{https://www.usenix.org/conference/osdi24/presentation/agrawal}
\showURL{%
\tempurl}


\bibitem[Ainslie et~al\mbox{.}(2023)]%
        {ainslie2023gqa}
\bibfield{author}{\bibinfo{person}{Joshua Ainslie}, \bibinfo{person}{James Lee-Thorp}, \bibinfo{person}{Michiel de Jong}, \bibinfo{person}{Yury Zemlyanskiy}, \bibinfo{person}{Federico Lebron}, {and} \bibinfo{person}{Sumit Sanghai}.} \bibinfo{year}{2023}\natexlab{}.
\newblock \showarticletitle{GQA: Training Generalized Multi-Query Transformer Models from Multi-Head Checkpoints}. In \bibinfo{booktitle}{\emph{Proceedings of the 2023 Conference on Empirical Methods in Natural Language Processing}}. \bibinfo{pages}{4895--4901}.
\newblock


\bibitem[An et~al\mbox{.}(2023)]%
        {an2023leval}
\bibfield{author}{\bibinfo{person}{Chenxin An}, \bibinfo{person}{Shansan Gong}, \bibinfo{person}{Ming Zhong}, \bibinfo{person}{Mukai Li}, \bibinfo{person}{Jun Zhang}, \bibinfo{person}{Lingpeng Kong}, {and} \bibinfo{person}{Xipeng Qiu}.} \bibinfo{year}{2023}\natexlab{}.
\newblock \bibinfo{title}{L-Eval: Instituting Standardized Evaluation for Long Context Language Models}.
\newblock
\newblock
\showeprint[arxiv]{2307.11088}~[cs.CL]


\bibitem[anthropic(2024)]%
        {claude}
\bibfield{author}{\bibinfo{person}{anthropic}.} \bibinfo{year}{2024}\natexlab{}.
\newblock \bibinfo{title}{Claude}.
\newblock \bibinfo{howpublished}{\url{https://claude.ai/}}.
\newblock
\newblock
\shownote{[Computer software]}.


\bibitem[authors(2024a)]%
        {lightllm}
\bibfield{author}{\bibinfo{person}{LightLLM authors}.} \bibinfo{year}{2024}\natexlab{a}.
\newblock \bibinfo{title}{LightLLM}.
\newblock \bibinfo{howpublished}{\url{https://github.com/ModelTC/lightllm}}.
\newblock
\newblock
\shownote{[Computer software]}.


\bibitem[authors(2024b)]%
        {rtp-llm}
\bibfield{author}{\bibinfo{person}{RTP-LLM authors}.} \bibinfo{year}{2024}\natexlab{b}.
\newblock \bibinfo{title}{RTP-LLM}.
\newblock \bibinfo{howpublished}{\url{https://github.com/alibaba/rtp-llm}}.
\newblock
\newblock
\shownote{[Computer software]}.


\bibitem[Bae et~al\mbox{.}(2021)]%
        {lee2021flashneuron}
\bibfield{author}{\bibinfo{person}{Jonghyun Bae}, \bibinfo{person}{Jongsung Lee}, \bibinfo{person}{Yunho Jin}, \bibinfo{person}{Sam Son}, \bibinfo{person}{Shine Kim}, \bibinfo{person}{Hakbeom Jang}, \bibinfo{person}{Tae~Jun Ham}, {and} \bibinfo{person}{Jae~W. Lee}.} \bibinfo{year}{2021}\natexlab{}.
\newblock \showarticletitle{FlashNeuron: SSD-Enabled Large-Batch Training of Very Deep Neural Networks}. In \bibinfo{booktitle}{\emph{19th {USENIX} Conference on File and Storage Technologies ({FAST} 21)}}. \bibinfo{publisher}{{USENIX} Association}, \bibinfo{pages}{387--401}.
\newblock
\showISBNx{978-1-939133-20-5}
\urldef\tempurl%
\url{https://www.usenix.org/conference/fast21/presentation/bae}
\showURL{%
\tempurl}


\bibitem[Borgeaud et~al\mbox{.}(2021)]%
        {Borgeaud2021ImprovingLM}
\bibfield{author}{\bibinfo{person}{Sebastian Borgeaud}, \bibinfo{person}{Arthur Mensch}, \bibinfo{person}{Jordan Hoffmann}, \bibinfo{person}{Trevor Cai}, \bibinfo{person}{Eliza Rutherford}, \bibinfo{person}{Katie Millican}, \bibinfo{person}{George van~den Driessche}, \bibinfo{person}{Jean-Baptiste Lespiau}, \bibinfo{person}{Bogdan Damoc}, \bibinfo{person}{Aidan Clark}, \bibinfo{person}{Diego de Las~Casas}, \bibinfo{person}{Aurelia Guy}, \bibinfo{person}{Jacob Menick}, \bibinfo{person}{Roman Ring}, \bibinfo{person}{T.~W. Hennigan}, \bibinfo{person}{Saffron Huang}, \bibinfo{person}{Lorenzo Maggiore}, \bibinfo{person}{Chris Jones}, \bibinfo{person}{Albin Cassirer}, \bibinfo{person}{Andy Brock}, \bibinfo{person}{Michela Paganini}, \bibinfo{person}{Geoffrey Irving}, \bibinfo{person}{Oriol Vinyals}, \bibinfo{person}{Simon Osindero}, \bibinfo{person}{Karen Simonyan}, \bibinfo{person}{Jack~W. Rae}, \bibinfo{person}{Erich Elsen}, {and} \bibinfo{person}{L. Sifre}.} \bibinfo{year}{2021}\natexlab{}.
\newblock \showarticletitle{Improving language models by retrieving from trillions of tokens}. In \bibinfo{booktitle}{\emph{International Conference on Machine Learning}}.
\newblock
\urldef\tempurl%
\url{https://api.semanticscholar.org/CorpusID:244954723}
\showURL{%
\tempurl}


\bibitem[Cao et~al\mbox{.}(2023)]%
        {cao2023comprehensive}
\bibfield{author}{\bibinfo{person}{Yihan Cao}, \bibinfo{person}{Siyu Li}, \bibinfo{person}{Yixin Liu}, \bibinfo{person}{Zhiling Yan}, \bibinfo{person}{Yutong Dai}, \bibinfo{person}{Philip~S Yu}, {and} \bibinfo{person}{Lichao Sun}.} \bibinfo{year}{2023}\natexlab{}.
\newblock \showarticletitle{A comprehensive survey of ai-generated content (aigc): A history of generative ai from gan to chatgpt}.
\newblock \bibinfo{journal}{\emph{arXiv preprint arXiv:2303.04226}} (\bibinfo{year}{2023}).
\newblock


\bibitem[Chase(2022)]%
        {Chase_LangChain_2022}
\bibfield{author}{\bibinfo{person}{Harrison Chase}.} \bibinfo{year}{2022}\natexlab{}.
\newblock \bibinfo{booktitle}{\emph{{LangChain}}}.
\newblock
\urldef\tempurl%
\url{https://github.com/langchain-ai/langchain}
\showURL{%
\tempurl}


\bibitem[Chen et~al\mbox{.}(2021)]%
        {chen2021evaluating}
\bibfield{author}{\bibinfo{person}{Mark Chen}, \bibinfo{person}{Jerry Tworek}, \bibinfo{person}{Heewoo Jun}, \bibinfo{person}{Qiming Yuan}, \bibinfo{person}{Henrique~Ponde de Oliveira~Pinto}, \bibinfo{person}{Jared Kaplan}, \bibinfo{person}{Harri Edwards}, \bibinfo{person}{Yuri Burda}, \bibinfo{person}{Nicholas Joseph}, \bibinfo{person}{Greg Brockman}, \bibinfo{person}{Alex Ray}, \bibinfo{person}{Raul Puri}, \bibinfo{person}{Gretchen Krueger}, \bibinfo{person}{Michael Petrov}, \bibinfo{person}{Heidy Khlaaf}, \bibinfo{person}{Girish Sastry}, \bibinfo{person}{Pamela Mishkin}, \bibinfo{person}{Brooke Chan}, \bibinfo{person}{Scott Gray}, \bibinfo{person}{Nick Ryder}, \bibinfo{person}{Mikhail Pavlov}, \bibinfo{person}{Alethea Power}, \bibinfo{person}{Lukasz Kaiser}, \bibinfo{person}{Mohammad Bavarian}, \bibinfo{person}{Clemens Winter}, \bibinfo{person}{Philippe Tillet}, \bibinfo{person}{Felipe~Petroski Such}, \bibinfo{person}{Dave Cummings}, \bibinfo{person}{Matthias Plappert}, \bibinfo{person}{Fotios Chantzis},
  \bibinfo{person}{Elizabeth Barnes}, \bibinfo{person}{Ariel Herbert-Voss}, \bibinfo{person}{William~Hebgen Guss}, \bibinfo{person}{Alex Nichol}, \bibinfo{person}{Alex Paino}, \bibinfo{person}{Nikolas Tezak}, \bibinfo{person}{Jie Tang}, \bibinfo{person}{Igor Babuschkin}, \bibinfo{person}{Suchir Balaji}, \bibinfo{person}{Shantanu Jain}, \bibinfo{person}{William Saunders}, \bibinfo{person}{Christopher Hesse}, \bibinfo{person}{Andrew~N. Carr}, \bibinfo{person}{Jan Leike}, \bibinfo{person}{Josh Achiam}, \bibinfo{person}{Vedant Misra}, \bibinfo{person}{Evan Morikawa}, \bibinfo{person}{Alec Radford}, \bibinfo{person}{Matthew Knight}, \bibinfo{person}{Miles Brundage}, \bibinfo{person}{Mira Murati}, \bibinfo{person}{Katie Mayer}, \bibinfo{person}{Peter Welinder}, \bibinfo{person}{Bob McGrew}, \bibinfo{person}{Dario Amodei}, \bibinfo{person}{Sam McCandlish}, \bibinfo{person}{Ilya Sutskever}, {and} \bibinfo{person}{Wojciech Zaremba}.} \bibinfo{year}{2021}\natexlab{}.
\newblock \bibinfo{title}{Evaluating Large Language Models Trained on Code}.
\newblock
\newblock
\showeprint[arxiv]{2107.03374}~[cs.LG]


\bibitem[Cobbe et~al\mbox{.}(2021)]%
        {cobbe2021training}
\bibfield{author}{\bibinfo{person}{Karl Cobbe}, \bibinfo{person}{Vineet Kosaraju}, \bibinfo{person}{Mohammad Bavarian}, \bibinfo{person}{Mark Chen}, \bibinfo{person}{Heewoo Jun}, \bibinfo{person}{Lukasz Kaiser}, \bibinfo{person}{Matthias Plappert}, \bibinfo{person}{Jerry Tworek}, \bibinfo{person}{Jacob Hilton}, \bibinfo{person}{Reiichiro Nakano}, {et~al\mbox{.}}} \bibinfo{year}{2021}\natexlab{}.
\newblock \showarticletitle{Training verifiers to solve math word problems}.
\newblock \bibinfo{journal}{\emph{arXiv preprint arXiv:2110.14168}} (\bibinfo{year}{2021}).
\newblock


\bibitem[DeepSpeed(2024)]%
        {MII}
\bibfield{author}{\bibinfo{person}{DeepSpeed}.} \bibinfo{year}{2024}\natexlab{}.
\newblock \bibinfo{title}{DeepSpeed-MII}.
\newblock \bibinfo{howpublished}{\url{https://github.com/microsoft/DeepSpeed-MII}}.
\newblock
\newblock
\shownote{[Computer software]}.


\bibitem[Dong et~al\mbox{.}(2024)]%
        {dong2024qaq}
\bibfield{author}{\bibinfo{person}{Shichen Dong}, \bibinfo{person}{Wen Cheng}, \bibinfo{person}{Jiayu Qin}, {and} \bibinfo{person}{Wei Wang}.} \bibinfo{year}{2024}\natexlab{}.
\newblock \showarticletitle{QAQ: Quality Adaptive Quantization for LLM KV Cache}.
\newblock \bibinfo{journal}{\emph{arXiv preprint arXiv:2403.04643}} (\bibinfo{year}{2024}).
\newblock


\bibitem[Feng et~al\mbox{.}(2023)]%
        {feng23mobius}
\bibfield{author}{\bibinfo{person}{Yangyang Feng}, \bibinfo{person}{Minhui Xie}, \bibinfo{person}{Zijie Tian}, \bibinfo{person}{Shuo Wang}, \bibinfo{person}{Youyou Lu}, {and} \bibinfo{person}{Jiwu Shu}.} \bibinfo{year}{2023}\natexlab{}.
\newblock \showarticletitle{Mobius: Fine Tuning Large-Scale Models on Commodity GPU Servers}. In \bibinfo{booktitle}{\emph{Proceedings of the 28th ACM International Conference on Architectural Support for Programming Languages and Operating Systems, Volume 2}} (Vancouver, BC, Canada) \emph{(\bibinfo{series}{ASPLOS 2023})}. \bibinfo{publisher}{Association for Computing Machinery}, \bibinfo{address}{New York, NY, USA}, \bibinfo{pages}{489–501}.
\newblock
\showISBNx{9781450399166}
\urldef\tempurl%
\url{https://doi.org/10.1145/3575693.3575703}
\showDOI{\tempurl}


\bibitem[Frantar et~al\mbox{.}(2022)]%
        {frantar2022optq}
\bibfield{author}{\bibinfo{person}{Elias Frantar}, \bibinfo{person}{Saleh Ashkboos}, \bibinfo{person}{Torsten Hoefler}, {and} \bibinfo{person}{Dan Alistarh}.} \bibinfo{year}{2022}\natexlab{}.
\newblock \showarticletitle{OPTQ: Accurate quantization for generative pre-trained transformers}. In \bibinfo{booktitle}{\emph{The Eleventh International Conference on Learning Representations}}.
\newblock


\bibitem[Gao et~al\mbox{.}(2024)]%
        {zuo2024as}
\bibfield{author}{\bibinfo{person}{Bin Gao}, \bibinfo{person}{Zhuomin He}, \bibinfo{person}{Puru Sharma}, \bibinfo{person}{Qingxuan Kang}, \bibinfo{person}{Djordje Jevdjic}, \bibinfo{person}{Junbo Deng}, \bibinfo{person}{Xingkun Yang}, \bibinfo{person}{Zhou Yu}, {and} \bibinfo{person}{Pengfei Zuo}.} \bibinfo{year}{2024}\natexlab{}.
\newblock \showarticletitle{{Cost-Efficient} Large Language Model Serving for Multi-turn Conversations with {CachedAttention}}. In \bibinfo{booktitle}{\emph{2024 USENIX Annual Technical Conference (USENIX ATC 24)}}. \bibinfo{publisher}{USENIX Association}, \bibinfo{address}{Santa Clara, CA}, \bibinfo{pages}{111--126}.
\newblock
\showISBNx{978-1-939133-41-0}
\urldef\tempurl%
\url{https://www.usenix.org/conference/atc24/presentation/gao-bin-cost}
\showURL{%
\tempurl}


\bibitem[Gim et~al\mbox{.}(2024)]%
        {gim2024prompt}
\bibfield{author}{\bibinfo{person}{In Gim}, \bibinfo{person}{Guojun Chen}, \bibinfo{person}{Seung-seob Lee}, \bibinfo{person}{Nikhil Sarda}, \bibinfo{person}{Anurag Khandelwal}, {and} \bibinfo{person}{Lin Zhong}.} \bibinfo{year}{2024}\natexlab{}.
\newblock \showarticletitle{Prompt Cache: Modular Attention Reuse for Low-Latency Inference}. In \bibinfo{booktitle}{\emph{Proceedings of Machine Learning and Systems}}, \bibfield{editor}{\bibinfo{person}{P.~Gibbons}, \bibinfo{person}{G.~Pekhimenko}, {and} \bibinfo{person}{C.~De Sa}} (Eds.), Vol.~\bibinfo{volume}{6}. \bibinfo{pages}{325--338}.
\newblock
\urldef\tempurl%
\url{https://proceedings.mlsys.org/paper_files/paper/2024/file/a66caa1703fe34705a4368c3014c1966-Paper-Conference.pdf}
\showURL{%
\tempurl}


\bibitem[Gray and Neuhoff(1998)]%
        {gray1998quantization}
\bibfield{author}{\bibinfo{person}{Robert~M. Gray} {and} \bibinfo{person}{David~L. Neuhoff}.} \bibinfo{year}{1998}\natexlab{}.
\newblock \showarticletitle{Quantization}.
\newblock \bibinfo{journal}{\emph{IEEE transactions on information theory}} \bibinfo{volume}{44}, \bibinfo{number}{6} (\bibinfo{year}{1998}), \bibinfo{pages}{2325--2383}.
\newblock


\bibitem[Holmes et~al\mbox{.}(2024)]%
        {deepspeed2024fastgen}
\bibfield{author}{\bibinfo{person}{Connor Holmes}, \bibinfo{person}{Masahiro Tanaka}, \bibinfo{person}{Michael Wyatt}, \bibinfo{person}{Ammar~Ahmad Awan}, \bibinfo{person}{Jeff Rasley}, \bibinfo{person}{Samyam Rajbhandari}, \bibinfo{person}{Reza~Yazdani Aminabadi}, \bibinfo{person}{Heyang Qin}, \bibinfo{person}{Arash Bakhtiari}, \bibinfo{person}{Lev Kurilenko}, {and} \bibinfo{person}{Yuxiong He}.} \bibinfo{year}{2024}\natexlab{}.
\newblock \bibinfo{title}{DeepSpeed-FastGen: High-throughput Text Generation for LLMs via MII and DeepSpeed-Inference}.
\newblock
\newblock
\showeprint[arxiv]{2401.08671}~[cs.PF]


\bibitem[Hooper et~al\mbox{.}(2024)]%
        {hooper2024kvquant}
\bibfield{author}{\bibinfo{person}{Coleman Hooper}, \bibinfo{person}{Sehoon Kim}, \bibinfo{person}{Hiva Mohammadzadeh}, \bibinfo{person}{Michael~W Mahoney}, \bibinfo{person}{Yakun~Sophia Shao}, \bibinfo{person}{Kurt Keutzer}, {and} \bibinfo{person}{Amir Gholami}.} \bibinfo{year}{2024}\natexlab{}.
\newblock \showarticletitle{KVQuant: Towards 10 Million Context Length LLM Inference with KV Cache Quantization}.
\newblock \bibinfo{journal}{\emph{arXiv preprint arXiv:2401.18079}} (\bibinfo{year}{2024}).
\newblock


\bibitem[Huang et~al\mbox{.}(2023)]%
        {huang2023survey}
\bibfield{author}{\bibinfo{person}{Lei Huang}, \bibinfo{person}{Weijiang Yu}, \bibinfo{person}{Weitao Ma}, \bibinfo{person}{Weihong Zhong}, \bibinfo{person}{Zhangyin Feng}, \bibinfo{person}{Haotian Wang}, \bibinfo{person}{Qianglong Chen}, \bibinfo{person}{Weihua Peng}, \bibinfo{person}{Xiaocheng Feng}, \bibinfo{person}{Bing Qin}, {and} \bibinfo{person}{Ting Liu}.} \bibinfo{year}{2023}\natexlab{}.
\newblock \bibinfo{title}{A Survey on Hallucination in Large Language Models: Principles, Taxonomy, Challenges, and Open Questions}.
\newblock
\newblock
\showeprint[arxiv]{2311.05232}~[cs.CL]


\bibitem[Javaheripi et~al\mbox{.}(2023)]%
        {javaheripi2023phi}
\bibfield{author}{\bibinfo{person}{Mojan Javaheripi}, \bibinfo{person}{S{\'e}bastien Bubeck}, \bibinfo{person}{Marah Abdin}, \bibinfo{person}{Jyoti Aneja}, \bibinfo{person}{Sebastien Bubeck}, \bibinfo{person}{Caio C{\'e}sar~Teodoro Mendes}, \bibinfo{person}{Weizhu Chen}, \bibinfo{person}{Allie Del~Giorno}, \bibinfo{person}{Ronen Eldan}, \bibinfo{person}{Sivakanth Gopi}, {et~al\mbox{.}}} \bibinfo{year}{2023}\natexlab{}.
\newblock \showarticletitle{Phi-2: The surprising power of small language models}.
\newblock \bibinfo{journal}{\emph{Microsoft Research Blog}} (\bibinfo{year}{2023}).
\newblock


\bibitem[Jin et~al\mbox{.}(2024)]%
        {jin2024ragcache}
\bibfield{author}{\bibinfo{person}{Chao Jin}, \bibinfo{person}{Zili Zhang}, \bibinfo{person}{Xuanlin Jiang}, \bibinfo{person}{Fangyue Liu}, \bibinfo{person}{Xin Liu}, \bibinfo{person}{Xuanzhe Liu}, {and} \bibinfo{person}{Xin Jin}.} \bibinfo{year}{2024}\natexlab{}.
\newblock \bibinfo{title}{RAGCache: Efficient Knowledge Caching for Retrieval-Augmented Generation}.
\newblock
\newblock
\showeprint[arxiv]{2404.12457}~[cs.DC]


\bibitem[Kang et~al\mbox{.}(2024)]%
        {kang2024gear}
\bibfield{author}{\bibinfo{person}{Hao Kang}, \bibinfo{person}{Qingru Zhang}, \bibinfo{person}{Souvik Kundu}, \bibinfo{person}{Geonhwa Jeong}, \bibinfo{person}{Zaoxing Liu}, \bibinfo{person}{Tushar Krishna}, {and} \bibinfo{person}{Tuo Zhao}.} \bibinfo{year}{2024}\natexlab{}.
\newblock \bibinfo{title}{GEAR: An Efficient KV Cache Compression Recipe for Near-Lossless Generative Inference of LLM}.
\newblock
\newblock
\showeprint[arxiv]{2403.05527}~[cs.LG]


\bibitem[Kwon et~al\mbox{.}(2023)]%
        {stoica2023vllm}
\bibfield{author}{\bibinfo{person}{Woosuk Kwon}, \bibinfo{person}{Zhuohan Li}, \bibinfo{person}{Siyuan Zhuang}, \bibinfo{person}{Ying Sheng}, \bibinfo{person}{Lianmin Zheng}, \bibinfo{person}{Cody~Hao Yu}, \bibinfo{person}{Joseph Gonzalez}, \bibinfo{person}{Hao Zhang}, {and} \bibinfo{person}{Ion Stoica}.} \bibinfo{year}{2023}\natexlab{}.
\newblock \showarticletitle{Efficient Memory Management for Large Language Model Serving with PagedAttention}. In \bibinfo{booktitle}{\emph{Proceedings of the 29th Symposium on Operating Systems Principles}} (, Koblenz, Germany,) \emph{(\bibinfo{series}{SOSP '23})}. \bibinfo{publisher}{Association for Computing Machinery}, \bibinfo{address}{New York, NY, USA}, \bibinfo{pages}{611–626}.
\newblock
\showISBNx{9798400702297}
\urldef\tempurl%
\url{https://doi.org/10.1145/3600006.3613165}
\showDOI{\tempurl}


\bibitem[Lewis et~al\mbox{.}(2020)]%
        {lewis2020retrieval}
\bibfield{author}{\bibinfo{person}{Patrick Lewis}, \bibinfo{person}{Ethan Perez}, \bibinfo{person}{Aleksandra Piktus}, \bibinfo{person}{Fabio Petroni}, \bibinfo{person}{Vladimir Karpukhin}, \bibinfo{person}{Naman Goyal}, \bibinfo{person}{Heinrich K{\"u}ttler}, \bibinfo{person}{Mike Lewis}, \bibinfo{person}{Wen-tau Yih}, \bibinfo{person}{Tim Rockt{\"a}schel}, {et~al\mbox{.}}} \bibinfo{year}{2020}\natexlab{}.
\newblock \showarticletitle{Retrieval-augmented generation for knowledge-intensive nlp tasks}.
\newblock \bibinfo{journal}{\emph{Advances in Neural Information Processing Systems}}  \bibinfo{volume}{33} (\bibinfo{year}{2020}), \bibinfo{pages}{9459--9474}.
\newblock


\bibitem[Lin et~al\mbox{.}(2023)]%
        {lin2023awq}
\bibfield{author}{\bibinfo{person}{Ji Lin}, \bibinfo{person}{Jiaming Tang}, \bibinfo{person}{Haotian Tang}, \bibinfo{person}{Shang Yang}, \bibinfo{person}{Xingyu Dang}, {and} \bibinfo{person}{Song Han}.} \bibinfo{year}{2023}\natexlab{}.
\newblock \showarticletitle{Awq: Activation-aware weight quantization for llm compression and acceleration}.
\newblock \bibinfo{journal}{\emph{arXiv preprint arXiv:2306.00978}} (\bibinfo{year}{2023}).
\newblock


\bibitem[Liu(2022)]%
        {Liu_LlamaIndex_2022}
\bibfield{author}{\bibinfo{person}{Jerry Liu}.} \bibinfo{year}{2022}\natexlab{}.
\newblock \bibinfo{booktitle}{\emph{{LlamaIndex}}}.
\newblock
\urldef\tempurl%
\url{https://doi.org/10.5281/zenodo.1234}
\showDOI{\tempurl}


\bibitem[Liu et~al\mbox{.}(2024a)]%
        {liu2024cachegen}
\bibfield{author}{\bibinfo{person}{Yuhan Liu}, \bibinfo{person}{Hanchen Li}, \bibinfo{person}{Yihua Cheng}, \bibinfo{person}{Siddhant Ray}, \bibinfo{person}{Yuyang Huang}, \bibinfo{person}{Qizheng Zhang}, \bibinfo{person}{Kuntai Du}, \bibinfo{person}{Jiayi Yao}, \bibinfo{person}{Shan Lu}, \bibinfo{person}{Ganesh Ananthanarayanan}, \bibinfo{person}{Michael Maire}, \bibinfo{person}{Henry Hoffmann}, \bibinfo{person}{Ari Holtzman}, {and} \bibinfo{person}{Junchen Jiang}.} \bibinfo{year}{2024}\natexlab{a}.
\newblock \showarticletitle{CacheGen: KV Cache Compression and Streaming for Fast Large Language Model Serving}. In \bibinfo{booktitle}{\emph{Proceedings of the ACM SIGCOMM 2024 Conference}} (Sydney, NSW, Australia) \emph{(\bibinfo{series}{ACM SIGCOMM '24})}. \bibinfo{publisher}{Association for Computing Machinery}, \bibinfo{address}{New York, NY, USA}, \bibinfo{pages}{38–56}.
\newblock
\showISBNx{9798400706141}
\urldef\tempurl%
\url{https://doi.org/10.1145/3651890.3672274}
\showDOI{\tempurl}


\bibitem[Liu et~al\mbox{.}(2024b)]%
        {liu2024kivi}
\bibfield{author}{\bibinfo{person}{Zirui Liu}, \bibinfo{person}{Jiayi Yuan}, \bibinfo{person}{Hongye Jin}, \bibinfo{person}{Shaochen Zhong}, \bibinfo{person}{Zhaozhuo Xu}, \bibinfo{person}{Vladimir Braverman}, \bibinfo{person}{Beidi Chen}, {and} \bibinfo{person}{Xia Hu}.} \bibinfo{year}{2024}\natexlab{b}.
\newblock \showarticletitle{KIVI: A Tuning-Free Asymmetric 2bit Quantization for KV Cache}.
\newblock \bibinfo{journal}{\emph{arXiv preprint arXiv:2402.02750}} (\bibinfo{year}{2024}).
\newblock


\bibitem[Mu et~al\mbox{.}(2023)]%
        {mu2023learning}
\bibfield{author}{\bibinfo{person}{Jesse Mu}, \bibinfo{person}{Xiang~Lisa Li}, {and} \bibinfo{person}{Noah Goodman}.} \bibinfo{year}{2023}\natexlab{}.
\newblock \showarticletitle{Learning to Compress Prompts with Gist Tokens}. In \bibinfo{booktitle}{\emph{Thirty-seventh Conference on Neural Information Processing Systems}}.
\newblock
\urldef\tempurl%
\url{https://openreview.net/forum?id=2DtxPCL3T5}
\showURL{%
\tempurl}


\bibitem[Nam et~al\mbox{.}(2024)]%
        {nam2024using}
\bibfield{author}{\bibinfo{person}{Daye Nam}, \bibinfo{person}{Andrew Macvean}, \bibinfo{person}{Vincent Hellendoorn}, \bibinfo{person}{Bogdan Vasilescu}, {and} \bibinfo{person}{Brad Myers}.} \bibinfo{year}{2024}\natexlab{}.
\newblock \showarticletitle{Using an llm to help with code understanding}. In \bibinfo{booktitle}{\emph{Proceedings of the IEEE/ACM 46th International Conference on Software Engineering}}. \bibinfo{pages}{1--13}.
\newblock


\bibitem[Narayanan et~al\mbox{.}(2019)]%
        {nara19pipedream}
\bibfield{author}{\bibinfo{person}{Deepak Narayanan}, \bibinfo{person}{Aaron Harlap}, \bibinfo{person}{Amar Phanishayee}, \bibinfo{person}{Vivek Seshadri}, \bibinfo{person}{Nikhil~R. Devanur}, \bibinfo{person}{Gregory~R. Ganger}, \bibinfo{person}{Phillip~B. Gibbons}, {and} \bibinfo{person}{Matei Zaharia}.} \bibinfo{year}{2019}\natexlab{}.
\newblock \showarticletitle{PipeDream: generalized pipeline parallelism for DNN training}. In \bibinfo{booktitle}{\emph{Proceedings of the 27th ACM Symposium on Operating Systems Principles}} (Huntsville, Ontario, Canada) \emph{(\bibinfo{series}{SOSP '19})}. \bibinfo{publisher}{Association for Computing Machinery}, \bibinfo{address}{New York, NY, USA}, \bibinfo{pages}{1–15}.
\newblock
\showISBNx{9781450368735}
\urldef\tempurl%
\url{https://doi.org/10.1145/3341301.3359646}
\showDOI{\tempurl}


\bibitem[NVIDIA(2024a)]%
        {gdrcopy}
\bibfield{author}{\bibinfo{person}{NVIDIA}.} \bibinfo{year}{2024}\natexlab{a}.
\newblock \bibinfo{title}{gdrcopy}.
\newblock \bibinfo{howpublished}{\url{https://github.com/NVIDIA/gdrcopy}}.
\newblock
\newblock
\shownote{[Computer software]}.


\bibitem[NVIDIA(2024b)]%
        {claudelong}
\bibfield{author}{\bibinfo{person}{NVIDIA}.} \bibinfo{year}{2024}\natexlab{b}.
\newblock \bibinfo{title}{Long context prompting for Claude 2.1}.
\newblock \bibinfo{howpublished}{\url{https://www.anthropic.com/news/claude-2-1-prompting}}.
\newblock
\newblock
\shownote{[Website]}.


\bibitem[NVIDIA(2024c)]%
        {TensorRT-LLM}
\bibfield{author}{\bibinfo{person}{NVIDIA}.} \bibinfo{year}{2024}\natexlab{c}.
\newblock \bibinfo{title}{TensorRT-LLM}.
\newblock \bibinfo{howpublished}{\url{https://github.com/NVIDIA/TensorRT-LLM}}.
\newblock
\newblock
\shownote{[Computer software]}.


\bibitem[OpenAI(2024)]%
        {chatgpt}
\bibfield{author}{\bibinfo{person}{OpenAI}.} \bibinfo{year}{2024}\natexlab{}.
\newblock \bibinfo{title}{ChatGPT}.
\newblock \bibinfo{howpublished}{\url{https://chat.openai.com/}}.
\newblock
\newblock
\shownote{[Computer software]}.


\bibitem[Pang et~al\mbox{.}(2022)]%
        {pang-etal-2022-quality}
\bibfield{author}{\bibinfo{person}{Richard~Yuanzhe Pang}, \bibinfo{person}{Alicia Parrish}, \bibinfo{person}{Nitish Joshi}, \bibinfo{person}{Nikita Nangia}, \bibinfo{person}{Jason Phang}, \bibinfo{person}{Angelica Chen}, \bibinfo{person}{Vishakh Padmakumar}, \bibinfo{person}{Johnny Ma}, \bibinfo{person}{Jana Thompson}, \bibinfo{person}{He He}, {and} \bibinfo{person}{Samuel Bowman}.} \bibinfo{year}{2022}\natexlab{}.
\newblock \showarticletitle{{Q}u{ALITY}: Question Answering with Long Input Texts, Yes!}. In \bibinfo{booktitle}{\emph{Proceedings of the 2022 Conference of the North American Chapter of the Association for Computational Linguistics: Human Language Technologies}}. \bibinfo{publisher}{Association for Computational Linguistics}, \bibinfo{address}{Seattle, United States}, \bibinfo{pages}{5336--5358}.
\newblock
\urldef\tempurl%
\url{https://aclanthology.org/2022.naacl-main.391}
\showURL{%
\tempurl}


\bibitem[Peng et~al\mbox{.}(2023)]%
        {peng2023rwkv}
\bibfield{author}{\bibinfo{person}{Bo Peng}, \bibinfo{person}{Eric Alcaide}, \bibinfo{person}{Quentin Anthony}, \bibinfo{person}{Alon Albalak}, \bibinfo{person}{Samuel Arcadinho}, \bibinfo{person}{Huanqi Cao}, \bibinfo{person}{Xin Cheng}, \bibinfo{person}{Michael Chung}, \bibinfo{person}{Matteo Grella}, \bibinfo{person}{Kranthi~Kiran GV}, {et~al\mbox{.}}} \bibinfo{year}{2023}\natexlab{}.
\newblock \showarticletitle{Rwkv: Reinventing rnns for the transformer era}.
\newblock \bibinfo{journal}{\emph{arXiv preprint arXiv:2305.13048}} (\bibinfo{year}{2023}).
\newblock


\bibitem[Shazeer(2019)]%
        {shazeer2019fast}
\bibfield{author}{\bibinfo{person}{Noam Shazeer}.} \bibinfo{year}{2019}\natexlab{}.
\newblock \showarticletitle{Fast transformer decoding: One write-head is all you need}.
\newblock \bibinfo{journal}{\emph{arXiv preprint arXiv:1911.02150}} (\bibinfo{year}{2019}).
\newblock


\bibitem[Sheng et~al\mbox{.}(2023)]%
        {sheng2024flexgen}
\bibfield{author}{\bibinfo{person}{Ying Sheng}, \bibinfo{person}{Lianmin Zheng}, \bibinfo{person}{Binhang Yuan}, \bibinfo{person}{Zhuohan Li}, \bibinfo{person}{Max Ryabinin}, \bibinfo{person}{Beidi Chen}, \bibinfo{person}{Percy Liang}, \bibinfo{person}{Christopher R\'{e}}, \bibinfo{person}{Ion Stoica}, {and} \bibinfo{person}{Ce Zhang}.} \bibinfo{year}{2023}\natexlab{}.
\newblock \showarticletitle{FlexGen: high-throughput generative inference of large language models with a single GPU}. In \bibinfo{booktitle}{\emph{Proceedings of the 40th International Conference on Machine Learning}} (, Honolulu, Hawaii, USA,) \emph{(\bibinfo{series}{ICML'23})}. \bibinfo{publisher}{JMLR.org}, Article \bibinfo{articleno}{1288}, \bibinfo{numpages}{23}~pages.
\newblock


\bibitem[Song et~al\mbox{.}(2023)]%
        {song2023ugcache}
\bibfield{author}{\bibinfo{person}{Xiaoniu Song}, \bibinfo{person}{Yiwen Zhang}, \bibinfo{person}{Rong Chen}, {and} \bibinfo{person}{Haibo Chen}.} \bibinfo{year}{2023}\natexlab{}.
\newblock \showarticletitle{UGACHE: A Unified GPU Cache for Embedding-based Deep Learning}. In \bibinfo{booktitle}{\emph{Proceedings of the 29th Symposium on Operating Systems Principles}} (Koblenz, Germany) \emph{(\bibinfo{series}{SOSP '23})}. \bibinfo{publisher}{Association for Computing Machinery}, \bibinfo{address}{New York, NY, USA}, \bibinfo{pages}{627–641}.
\newblock
\showISBNx{9798400702297}
\urldef\tempurl%
\url{https://doi.org/10.1145/3600006.3613169}
\showDOI{\tempurl}


\bibitem[Su et~al\mbox{.}(2024)]%
        {su2024roformer}
\bibfield{author}{\bibinfo{person}{Jianlin Su}, \bibinfo{person}{Murtadha Ahmed}, \bibinfo{person}{Yu Lu}, \bibinfo{person}{Shengfeng Pan}, \bibinfo{person}{Wen Bo}, {and} \bibinfo{person}{Yunfeng Liu}.} \bibinfo{year}{2024}\natexlab{}.
\newblock \showarticletitle{Roformer: Enhanced transformer with rotary position embedding}.
\newblock \bibinfo{journal}{\emph{Neurocomputing}}  \bibinfo{volume}{568} (\bibinfo{year}{2024}), \bibinfo{pages}{127063}.
\newblock


\bibitem[Team et~al\mbox{.}(2024)]%
        {gemmateam2024gemma}
\bibfield{author}{\bibinfo{person}{Gemma Team}, \bibinfo{person}{Thomas Mesnard}, \bibinfo{person}{Cassidy Hardin}, \bibinfo{person}{Robert Dadashi}, \bibinfo{person}{Surya Bhupatiraju}, \bibinfo{person}{Shreya Pathak}, \bibinfo{person}{Laurent Sifre}, \bibinfo{person}{Morgane Rivière}, \bibinfo{person}{Mihir~Sanjay Kale}, \bibinfo{person}{Juliette Love}, \bibinfo{person}{Pouya Tafti}, \bibinfo{person}{Léonard Hussenot}, \bibinfo{person}{Pier~Giuseppe Sessa}, \bibinfo{person}{Aakanksha Chowdhery}, \bibinfo{person}{Adam Roberts}, \bibinfo{person}{Aditya Barua}, \bibinfo{person}{Alex Botev}, \bibinfo{person}{Alex Castro-Ros}, \bibinfo{person}{Ambrose Slone}, \bibinfo{person}{Amélie Héliou}, \bibinfo{person}{Andrea Tacchetti}, \bibinfo{person}{Anna Bulanova}, \bibinfo{person}{Antonia Paterson}, \bibinfo{person}{Beth Tsai}, \bibinfo{person}{Bobak Shahriari}, \bibinfo{person}{Charline~Le Lan}, \bibinfo{person}{Christopher~A. Choquette-Choo}, \bibinfo{person}{Clément Crepy}, \bibinfo{person}{Daniel Cer},
  \bibinfo{person}{Daphne Ippolito}, \bibinfo{person}{David Reid}, \bibinfo{person}{Elena Buchatskaya}, \bibinfo{person}{Eric Ni}, \bibinfo{person}{Eric Noland}, \bibinfo{person}{Geng Yan}, \bibinfo{person}{George Tucker}, \bibinfo{person}{George-Christian Muraru}, \bibinfo{person}{Grigory Rozhdestvenskiy}, \bibinfo{person}{Henryk Michalewski}, \bibinfo{person}{Ian Tenney}, \bibinfo{person}{Ivan Grishchenko}, \bibinfo{person}{Jacob Austin}, \bibinfo{person}{James Keeling}, \bibinfo{person}{Jane Labanowski}, \bibinfo{person}{Jean-Baptiste Lespiau}, \bibinfo{person}{Jeff Stanway}, \bibinfo{person}{Jenny Brennan}, \bibinfo{person}{Jeremy Chen}, \bibinfo{person}{Johan Ferret}, \bibinfo{person}{Justin Chiu}, \bibinfo{person}{Justin Mao-Jones}, \bibinfo{person}{Katherine Lee}, \bibinfo{person}{Kathy Yu}, \bibinfo{person}{Katie Millican}, \bibinfo{person}{Lars~Lowe Sjoesund}, \bibinfo{person}{Lisa Lee}, \bibinfo{person}{Lucas Dixon}, \bibinfo{person}{Machel Reid}, \bibinfo{person}{Maciej Mikuła},
  \bibinfo{person}{Mateo Wirth}, \bibinfo{person}{Michael Sharman}, \bibinfo{person}{Nikolai Chinaev}, \bibinfo{person}{Nithum Thain}, \bibinfo{person}{Olivier Bachem}, \bibinfo{person}{Oscar Chang}, \bibinfo{person}{Oscar Wahltinez}, \bibinfo{person}{Paige Bailey}, \bibinfo{person}{Paul Michel}, \bibinfo{person}{Petko Yotov}, \bibinfo{person}{Rahma Chaabouni}, \bibinfo{person}{Ramona Comanescu}, \bibinfo{person}{Reena Jana}, \bibinfo{person}{Rohan Anil}, \bibinfo{person}{Ross McIlroy}, \bibinfo{person}{Ruibo Liu}, \bibinfo{person}{Ryan Mullins}, \bibinfo{person}{Samuel~L Smith}, \bibinfo{person}{Sebastian Borgeaud}, \bibinfo{person}{Sertan Girgin}, \bibinfo{person}{Sholto Douglas}, \bibinfo{person}{Shree Pandya}, \bibinfo{person}{Siamak Shakeri}, \bibinfo{person}{Soham De}, \bibinfo{person}{Ted Klimenko}, \bibinfo{person}{Tom Hennigan}, \bibinfo{person}{Vlad Feinberg}, \bibinfo{person}{Wojciech Stokowiec}, \bibinfo{person}{Yu hui Chen}, \bibinfo{person}{Zafarali Ahmed}, \bibinfo{person}{Zhitao Gong},
  \bibinfo{person}{Tris Warkentin}, \bibinfo{person}{Ludovic Peran}, \bibinfo{person}{Minh Giang}, \bibinfo{person}{Clément Farabet}, \bibinfo{person}{Oriol Vinyals}, \bibinfo{person}{Jeff Dean}, \bibinfo{person}{Koray Kavukcuoglu}, \bibinfo{person}{Demis Hassabis}, \bibinfo{person}{Zoubin Ghahramani}, \bibinfo{person}{Douglas Eck}, \bibinfo{person}{Joelle Barral}, \bibinfo{person}{Fernando Pereira}, \bibinfo{person}{Eli Collins}, \bibinfo{person}{Armand Joulin}, \bibinfo{person}{Noah Fiedel}, \bibinfo{person}{Evan Senter}, \bibinfo{person}{Alek Andreev}, {and} \bibinfo{person}{Kathleen Kenealy}.} \bibinfo{year}{2024}\natexlab{}.
\newblock \bibinfo{title}{Gemma: Open Models Based on Gemini Research and Technology}.
\newblock
\newblock
\showeprint[arxiv]{2403.08295}~[cs.CL]


\bibitem[Tonmoy et~al\mbox{.}(2024)]%
        {tonmoy2024comprehensive}
\bibfield{author}{\bibinfo{person}{SM Tonmoy}, \bibinfo{person}{SM Zaman}, \bibinfo{person}{Vinija Jain}, \bibinfo{person}{Anku Rani}, \bibinfo{person}{Vipula Rawte}, \bibinfo{person}{Aman Chadha}, {and} \bibinfo{person}{Amitava Das}.} \bibinfo{year}{2024}\natexlab{}.
\newblock \showarticletitle{A comprehensive survey of hallucination mitigation techniques in large language models}.
\newblock \bibinfo{journal}{\emph{arXiv preprint arXiv:2401.01313}} (\bibinfo{year}{2024}).
\newblock


\bibitem[Touvron et~al\mbox{.}(2023)]%
        {touvron2023llama}
\bibfield{author}{\bibinfo{person}{Hugo Touvron}, \bibinfo{person}{Louis Martin}, \bibinfo{person}{Kevin Stone}, \bibinfo{person}{Peter Albert}, \bibinfo{person}{Amjad Almahairi}, \bibinfo{person}{Yasmine Babaei}, \bibinfo{person}{Nikolay Bashlykov}, \bibinfo{person}{Soumya Batra}, \bibinfo{person}{Prajjwal Bhargava}, \bibinfo{person}{Shruti Bhosale}, \bibinfo{person}{Dan Bikel}, \bibinfo{person}{Lukas Blecher}, \bibinfo{person}{Cristian~Canton Ferrer}, \bibinfo{person}{Moya Chen}, \bibinfo{person}{Guillem Cucurull}, \bibinfo{person}{David Esiobu}, \bibinfo{person}{Jude Fernandes}, \bibinfo{person}{Jeremy Fu}, \bibinfo{person}{Wenyin Fu}, \bibinfo{person}{Brian Fuller}, \bibinfo{person}{Cynthia Gao}, \bibinfo{person}{Vedanuj Goswami}, \bibinfo{person}{Naman Goyal}, \bibinfo{person}{Anthony Hartshorn}, \bibinfo{person}{Saghar Hosseini}, \bibinfo{person}{Rui Hou}, \bibinfo{person}{Hakan Inan}, \bibinfo{person}{Marcin Kardas}, \bibinfo{person}{Viktor Kerkez}, \bibinfo{person}{Madian Khabsa},
  \bibinfo{person}{Isabel Kloumann}, \bibinfo{person}{Artem Korenev}, \bibinfo{person}{Punit~Singh Koura}, \bibinfo{person}{Marie-Anne Lachaux}, \bibinfo{person}{Thibaut Lavril}, \bibinfo{person}{Jenya Lee}, \bibinfo{person}{Diana Liskovich}, \bibinfo{person}{Yinghai Lu}, \bibinfo{person}{Yuning Mao}, \bibinfo{person}{Xavier Martinet}, \bibinfo{person}{Todor Mihaylov}, \bibinfo{person}{Pushkar Mishra}, \bibinfo{person}{Igor Molybog}, \bibinfo{person}{Yixin Nie}, \bibinfo{person}{Andrew Poulton}, \bibinfo{person}{Jeremy Reizenstein}, \bibinfo{person}{Rashi Rungta}, \bibinfo{person}{Kalyan Saladi}, \bibinfo{person}{Alan Schelten}, \bibinfo{person}{Ruan Silva}, \bibinfo{person}{Eric~Michael Smith}, \bibinfo{person}{Ranjan Subramanian}, \bibinfo{person}{Xiaoqing~Ellen Tan}, \bibinfo{person}{Binh Tang}, \bibinfo{person}{Ross Taylor}, \bibinfo{person}{Adina Williams}, \bibinfo{person}{Jian~Xiang Kuan}, \bibinfo{person}{Puxin Xu}, \bibinfo{person}{Zheng Yan}, \bibinfo{person}{Iliyan Zarov}, \bibinfo{person}{Yuchen
  Zhang}, \bibinfo{person}{Angela Fan}, \bibinfo{person}{Melanie Kambadur}, \bibinfo{person}{Sharan Narang}, \bibinfo{person}{Aurelien Rodriguez}, \bibinfo{person}{Robert Stojnic}, \bibinfo{person}{Sergey Edunov}, {and} \bibinfo{person}{Thomas Scialom}.} \bibinfo{year}{2023}\natexlab{}.
\newblock \bibinfo{title}{Llama 2: Open Foundation and Fine-Tuned Chat Models}.
\newblock
\newblock
\showeprint[arxiv]{2307.09288}~[cs.CL]


\bibitem[Vaswani et~al\mbox{.}(2017)]%
        {vaswani2017attention}
\bibfield{author}{\bibinfo{person}{Ashish Vaswani}, \bibinfo{person}{Noam Shazeer}, \bibinfo{person}{Niki Parmar}, \bibinfo{person}{Jakob Uszkoreit}, \bibinfo{person}{Llion Jones}, \bibinfo{person}{Aidan~N Gomez}, \bibinfo{person}{{\L}ukasz Kaiser}, {and} \bibinfo{person}{Illia Polosukhin}.} \bibinfo{year}{2017}\natexlab{}.
\newblock \showarticletitle{Attention is all you need}.
\newblock \bibinfo{journal}{\emph{Advances in neural information processing systems}}  \bibinfo{volume}{30} (\bibinfo{year}{2017}).
\newblock


\bibitem[Wang et~al\mbox{.}(2024)]%
        {Wang_2024}
\bibfield{author}{\bibinfo{person}{Lei Wang}, \bibinfo{person}{Chen Ma}, \bibinfo{person}{Xueyang Feng}, \bibinfo{person}{Zeyu Zhang}, \bibinfo{person}{Hao Yang}, \bibinfo{person}{Jingsen Zhang}, \bibinfo{person}{Zhiyuan Chen}, \bibinfo{person}{Jiakai Tang}, \bibinfo{person}{Xu Chen}, \bibinfo{person}{Yankai Lin}, \bibinfo{person}{Wayne~Xin Zhao}, \bibinfo{person}{Zhewei Wei}, {and} \bibinfo{person}{Jirong Wen}.} \bibinfo{year}{2024}\natexlab{}.
\newblock \showarticletitle{A survey on large language model based autonomous agents}.
\newblock \bibinfo{journal}{\emph{Frontiers of Computer Science}} \bibinfo{volume}{18}, \bibinfo{number}{6} (\bibinfo{date}{March} \bibinfo{year}{2024}).
\newblock
\showISSN{2095-2236}
\urldef\tempurl%
\url{https://doi.org/10.1007/s11704-024-40231-1}
\showDOI{\tempurl}


\bibitem[Wu et~al\mbox{.}(2023)]%
        {wu2023fast}
\bibfield{author}{\bibinfo{person}{Bingyang Wu}, \bibinfo{person}{Yinmin Zhong}, \bibinfo{person}{Zili Zhang}, \bibinfo{person}{Gang Huang}, \bibinfo{person}{Xuanzhe Liu}, {and} \bibinfo{person}{Xin Jin}.} \bibinfo{year}{2023}\natexlab{}.
\newblock \showarticletitle{Fast distributed inference serving for large language models}.
\newblock \bibinfo{journal}{\emph{arXiv preprint arXiv:2305.05920}} (\bibinfo{year}{2023}).
\newblock


\bibitem[Xi et~al\mbox{.}(2023)]%
        {xi2023rise}
\bibfield{author}{\bibinfo{person}{Zhiheng Xi}, \bibinfo{person}{Wenxiang Chen}, \bibinfo{person}{Xin Guo}, \bibinfo{person}{Wei He}, \bibinfo{person}{Yiwen Ding}, \bibinfo{person}{Boyang Hong}, \bibinfo{person}{Ming Zhang}, \bibinfo{person}{Junzhe Wang}, \bibinfo{person}{Senjie Jin}, \bibinfo{person}{Enyu Zhou}, \bibinfo{person}{Rui Zheng}, \bibinfo{person}{Xiaoran Fan}, \bibinfo{person}{Xiao Wang}, \bibinfo{person}{Limao Xiong}, \bibinfo{person}{Yuhao Zhou}, \bibinfo{person}{Weiran Wang}, \bibinfo{person}{Changhao Jiang}, \bibinfo{person}{Yicheng Zou}, \bibinfo{person}{Xiangyang Liu}, \bibinfo{person}{Zhangyue Yin}, \bibinfo{person}{Shihan Dou}, \bibinfo{person}{Rongxiang Weng}, \bibinfo{person}{Wensen Cheng}, \bibinfo{person}{Qi Zhang}, \bibinfo{person}{Wenjuan Qin}, \bibinfo{person}{Yongyan Zheng}, \bibinfo{person}{Xipeng Qiu}, \bibinfo{person}{Xuanjing Huang}, {and} \bibinfo{person}{Tao Gui}.} \bibinfo{year}{2023}\natexlab{}.
\newblock \bibinfo{title}{The Rise and Potential of Large Language Model Based Agents: A Survey}.
\newblock
\newblock
\showeprint[arxiv]{2309.07864}~[cs.AI]


\bibitem[Xiao et~al\mbox{.}(2023)]%
        {pmlr-v202-xiao23c}
\bibfield{author}{\bibinfo{person}{Guangxuan Xiao}, \bibinfo{person}{Ji Lin}, \bibinfo{person}{Mickael Seznec}, \bibinfo{person}{Hao Wu}, \bibinfo{person}{Julien Demouth}, {and} \bibinfo{person}{Song Han}.} \bibinfo{year}{2023}\natexlab{}.
\newblock \showarticletitle{{S}mooth{Q}uant: Accurate and Efficient Post-Training Quantization for Large Language Models}. In \bibinfo{booktitle}{\emph{Proceedings of the 40th International Conference on Machine Learning}} \emph{(\bibinfo{series}{Proceedings of Machine Learning Research}, Vol.~\bibinfo{volume}{202})}, \bibfield{editor}{\bibinfo{person}{Andreas Krause}, \bibinfo{person}{Emma Brunskill}, \bibinfo{person}{Kyunghyun Cho}, \bibinfo{person}{Barbara Engelhardt}, \bibinfo{person}{Sivan Sabato}, {and} \bibinfo{person}{Jonathan Scarlett}} (Eds.). \bibinfo{publisher}{PMLR}, \bibinfo{pages}{38087--38099}.
\newblock
\urldef\tempurl%
\url{https://proceedings.mlr.press/v202/xiao23c.html}
\showURL{%
\tempurl}


\bibitem[Xie et~al\mbox{.}(2022)]%
        {xie2022fleche}
\bibfield{author}{\bibinfo{person}{Minhui Xie}, \bibinfo{person}{Youyou Lu}, \bibinfo{person}{Jiazhen Lin}, \bibinfo{person}{Qing Wang}, \bibinfo{person}{Jian Gao}, \bibinfo{person}{Kai Ren}, {and} \bibinfo{person}{Jiwu Shu}.} \bibinfo{year}{2022}\natexlab{}.
\newblock \showarticletitle{Fleche: an efficient GPU embedding cache for personalized recommendations}. In \bibinfo{booktitle}{\emph{Proceedings of the Seventeenth European Conference on Computer Systems}} (Rennes, France) \emph{(\bibinfo{series}{EuroSys '22})}. \bibinfo{publisher}{Association for Computing Machinery}, \bibinfo{address}{New York, NY, USA}, \bibinfo{pages}{402–416}.
\newblock
\showISBNx{9781450391627}
\urldef\tempurl%
\url{https://doi.org/10.1145/3492321.3519554}
\showDOI{\tempurl}


\bibitem[Xie et~al\mbox{.}(2023)]%
        {xie2023petps}
\bibfield{author}{\bibinfo{person}{Minhui Xie}, \bibinfo{person}{Youyou Lu}, \bibinfo{person}{Qing Wang}, \bibinfo{person}{Yangyang Feng}, \bibinfo{person}{Jiaqiang Liu}, \bibinfo{person}{Kai Ren}, {and} \bibinfo{person}{Jiwu Shu}.} \bibinfo{year}{2023}\natexlab{}.
\newblock \showarticletitle{PetPS: Supporting huge embedding models with persistent memory}.
\newblock \bibinfo{journal}{\emph{Proceedings of the VLDB Endowment}} \bibinfo{volume}{16}, \bibinfo{number}{5} (\bibinfo{year}{2023}), \bibinfo{pages}{1013--1022}.
\newblock


\bibitem[Xie et~al\mbox{.}(2020)]%
        {xie2020kraken}
\bibfield{author}{\bibinfo{person}{Minhui Xie}, \bibinfo{person}{Kai Ren}, \bibinfo{person}{Youyou Lu}, \bibinfo{person}{Guangxu Yang}, \bibinfo{person}{Qingxing Xu}, \bibinfo{person}{Bihai Wu}, \bibinfo{person}{Jiazhen Lin}, \bibinfo{person}{Hongbo Ao}, \bibinfo{person}{Wanhong Xu}, {and} \bibinfo{person}{Jiwu Shu}.} \bibinfo{year}{2020}\natexlab{}.
\newblock \showarticletitle{Kraken: memory-efficient continual learning for large-scale real-time recommendations}. In \bibinfo{booktitle}{\emph{SC20: International Conference for High Performance Computing, Networking, Storage and Analysis}}. IEEE, \bibinfo{pages}{1--17}.
\newblock


\bibitem[Yang et~al\mbox{.}(2024)]%
        {yang2024no}
\bibfield{author}{\bibinfo{person}{June~Yong Yang}, \bibinfo{person}{Byeongwook Kim}, \bibinfo{person}{Jeongin Bae}, \bibinfo{person}{Beomseok Kwon}, \bibinfo{person}{Gunho Park}, \bibinfo{person}{Eunho Yang}, \bibinfo{person}{Se~Jung Kwon}, {and} \bibinfo{person}{Dongsoo Lee}.} \bibinfo{year}{2024}\natexlab{}.
\newblock \showarticletitle{No Token Left Behind: Reliable KV Cache Compression via Importance-Aware Mixed Precision Quantization}.
\newblock \bibinfo{journal}{\emph{arXiv preprint arXiv:2402.18096}} (\bibinfo{year}{2024}).
\newblock


\bibitem[Yang et~al\mbox{.}(2017)]%
        {yang2017spdk}
\bibfield{author}{\bibinfo{person}{Ziye Yang}, \bibinfo{person}{James~R Harris}, \bibinfo{person}{Benjamin Walker}, \bibinfo{person}{Daniel Verkamp}, \bibinfo{person}{Changpeng Liu}, \bibinfo{person}{Cunyin Chang}, \bibinfo{person}{Gang Cao}, \bibinfo{person}{Jonathan Stern}, \bibinfo{person}{Vishal Verma}, {and} \bibinfo{person}{Luse~E Paul}.} \bibinfo{year}{2017}\natexlab{}.
\newblock \showarticletitle{SPDK: A development kit to build high performance storage applications}. In \bibinfo{booktitle}{\emph{2017 IEEE International Conference on Cloud Computing Technology and Science (CloudCom)}}. IEEE, \bibinfo{pages}{154--161}.
\newblock


\bibitem[Ye et~al\mbox{.}(2024)]%
        {ye2024chunkattention}
\bibfield{author}{\bibinfo{person}{Lu Ye}, \bibinfo{person}{Ze Tao}, \bibinfo{person}{Yong Huang}, {and} \bibinfo{person}{Yang Li}.} \bibinfo{year}{2024}\natexlab{}.
\newblock \bibinfo{title}{ChunkAttention: Efficient Attention on {KV} Cache with Chunking Sharing and Batching}.
\newblock
\newblock
\urldef\tempurl%
\url{https://openreview.net/forum?id=9k27IITeAZ}
\showURL{%
\tempurl}


\bibitem[Yu et~al\mbox{.}(2022)]%
        {yu2022orca}
\bibfield{author}{\bibinfo{person}{Gyeong-In Yu}, \bibinfo{person}{Joo~Seong Jeong}, \bibinfo{person}{Geon-Woo Kim}, \bibinfo{person}{Soojeong Kim}, {and} \bibinfo{person}{Byung-Gon Chun}.} \bibinfo{year}{2022}\natexlab{}.
\newblock \showarticletitle{Orca: A distributed serving system for $\{$Transformer-Based$\}$ generative models}. In \bibinfo{booktitle}{\emph{16th USENIX Symposium on Operating Systems Design and Implementation (OSDI 22)}}. \bibinfo{pages}{521--538}.
\newblock


\bibitem[Yu and Li(2023)]%
        {yu2023stateful}
\bibfield{author}{\bibinfo{person}{Lingfan Yu} {and} \bibinfo{person}{Jinyang Li}.} \bibinfo{year}{2023}\natexlab{}.
\newblock \bibinfo{title}{Stateful Large Language Model Serving with Pensieve}.
\newblock
\newblock
\showeprint[arxiv]{2312.05516}~[cs.LG]


\bibitem[Zhang et~al\mbox{.}(2024b)]%
        {zhang2024kv}
\bibfield{author}{\bibinfo{person}{Tianyi Zhang}, \bibinfo{person}{Jonah Yi}, \bibinfo{person}{Zhaozhuo Xu}, {and} \bibinfo{person}{Anshumali Shrivastava}.} \bibinfo{year}{2024}\natexlab{b}.
\newblock \showarticletitle{KV Cache is 1 Bit Per Channel: Efficient Large Language Model Inference with Coupled Quantization}.
\newblock \bibinfo{journal}{\emph{arXiv preprint arXiv:2405.03917}} (\bibinfo{year}{2024}).
\newblock


\bibitem[Zhang et~al\mbox{.}(2024a)]%
        {zhang2024h2o}
\bibfield{author}{\bibinfo{person}{Zhenyu Zhang}, \bibinfo{person}{Ying Sheng}, \bibinfo{person}{Tianyi Zhou}, \bibinfo{person}{Tianlong Chen}, \bibinfo{person}{Lianmin Zheng}, \bibinfo{person}{Ruisi Cai}, \bibinfo{person}{Zhao Song}, \bibinfo{person}{Yuandong Tian}, \bibinfo{person}{Christopher R{\'e}}, \bibinfo{person}{Clark Barrett}, {et~al\mbox{.}}} \bibinfo{year}{2024}\natexlab{a}.
\newblock \showarticletitle{H2o: Heavy-hitter oracle for efficient generative inference of large language models}.
\newblock \bibinfo{journal}{\emph{Advances in Neural Information Processing Systems}}  \bibinfo{volume}{36} (\bibinfo{year}{2024}).
\newblock


\bibitem[Zhao et~al\mbox{.}(2023a)]%
        {zhao2023atom}
\bibfield{author}{\bibinfo{person}{Yilong Zhao}, \bibinfo{person}{Chien-Yu Lin}, \bibinfo{person}{Kan Zhu}, \bibinfo{person}{Zihao Ye}, \bibinfo{person}{Lequn Chen}, \bibinfo{person}{Size Zheng}, \bibinfo{person}{Luis Ceze}, \bibinfo{person}{Arvind Krishnamurthy}, \bibinfo{person}{Tianqi Chen}, {and} \bibinfo{person}{Baris Kasikci}.} \bibinfo{year}{2023}\natexlab{a}.
\newblock \showarticletitle{Atom: Low-bit quantization for efficient and accurate llm serving}.
\newblock \bibinfo{journal}{\emph{arXiv preprint arXiv:2310.19102}} (\bibinfo{year}{2023}).
\newblock


\bibitem[Zhao et~al\mbox{.}(2023b)]%
        {zhao2023more}
\bibfield{author}{\bibinfo{person}{Zoie Zhao}, \bibinfo{person}{Sophie Song}, \bibinfo{person}{Bridget Duah}, \bibinfo{person}{Jamie Macbeth}, \bibinfo{person}{Scott Carter}, \bibinfo{person}{Monica~P Van}, \bibinfo{person}{Nayeli~Suseth Bravo}, \bibinfo{person}{Matthew Klenk}, \bibinfo{person}{Kate Sick}, {and} \bibinfo{person}{Alexandre~LS Filipowicz}.} \bibinfo{year}{2023}\natexlab{b}.
\newblock \showarticletitle{More human than human: LLM-generated narratives outperform human-LLM interleaved narratives}. In \bibinfo{booktitle}{\emph{Proceedings of the 15th Conference on Creativity and Cognition}}. \bibinfo{pages}{368--370}.
\newblock


\bibitem[Zheng et~al\mbox{.}(2023)]%
        {zheng2023efficiently}
\bibfield{author}{\bibinfo{person}{Lianmin Zheng}, \bibinfo{person}{Liangsheng Yin}, \bibinfo{person}{Zhiqiang Xie}, \bibinfo{person}{Jeff Huang}, \bibinfo{person}{Chuyue Sun}, \bibinfo{person}{Cody~Hao Yu}, \bibinfo{person}{Shiyi Cao}, \bibinfo{person}{Christos Kozyrakis}, \bibinfo{person}{Ion Stoica}, \bibinfo{person}{Joseph~E. Gonzalez}, \bibinfo{person}{Clark Barrett}, {and} \bibinfo{person}{Ying Sheng}.} \bibinfo{year}{2023}\natexlab{}.
\newblock \bibinfo{title}{Efficiently Programming Large Language Models using SGLang}.
\newblock
\newblock
\showeprint[arxiv]{2312.07104}~[cs.AI]


\end{thebibliography}
\end{document}